# Generating gapless land surface temperature with a high spatio-temporal resolution by fusing multi-source satellite-observed and model-simulated data

Jun Ma[a], Huanfeng Shen[a*], Penghai Wu[b], Jingan Wu[c], Meiling Gao[d], Chunlei Meng[e]

[a] *School of Resource and Environmental Sciences, Wuhan University, Wuhan 430079, China*
[b] *Information Materials and Intelligent Sensing Laboratory of Anhui Province, Anhui University, Hefei 230601, China*
[c] *School of Geospatial Engineering and Science, Sun Yat-sen University, Zhuhai 519082, China*
[d] *College of Geological Engineering and Geomatics, Chang'an University, Xi'an 710054, China*
[e] *Institute of Urban Meteorology, China Meteorological Administration, Beijing 100089, China*

**\*Corresponding author.**
*E-mail address:* shenhf@whu.edu.cn (H. Shen)

**Abstract:** Land surface temperature (LST) is a key parameter when monitoring land surface processes. However, cloud contamination and the tradeoff between the spatial and temporal resolutions greatly impede the access to high-quality thermal infrared (TIR) remote sensing data. Despite the massive efforts made to solve these dilemmas, it is still difficult to generate LST estimates with concurrent spatial completeness and a high spatio-temporal resolution. Land surface models (LSMs) can be used to simulate gapless LST with a high temporal resolution, but this usually comes with a low spatial resolution. In this paper, we present an integrated temperature fusion framework for satellite-observed and LSM-simulated LST data to map gapless LST at a 60-m spatial resolution and half-hourly temporal resolution. The global linear model (GloLM) model and the diurnal land surface temperature cycle (DTC) model are respectively performed as preprocessing steps for sensor and temporal normalization between the different LST data. The Landsat LST, Moderate Resolution Imaging Spectroradiometer (MODIS) LST, and Community Land Model Version 5.0 (CLM 5.0)-simulated LST are then fused using a filter-based spatio-temporal integrated fusion model. Evaluations were implemented in an urban-dominated region (the city of Wuhan in China) and a natural-dominated region (the Heihe River Basin in China), in terms of accuracy, spatial variability, and diurnal temporal dynamics. Results indicate that the fused LST is highly consistent with actual Landsat LST data (in situ LST measurements), in terms of a Pearson correlation coefficient of 0.94 (0.97–0.99), a mean absolute error of 0.71–0.98 K (0.82–3.17 K), and a root-mean-square error of 0.97–1.26 K (1.09–3.97 K). The generated diurnal Landsat-like LSTs under all weather conditions are able to support diurnal dynamic studies that are the most relevant to human activities, such as the study of urban heat islands (UHIs) and water resource management at the field scale.

1. **Introduction**

Land surface temperature (LST) exerts an important role in the land-atmosphere energy budget, water and carbon cycles, and energy fluxes at field, regional, and global scales (Anderson et al. 2012; Bojinski et al. 2014; Hansen et al. 2010; Li et al. 2013). Diurnal fine-resolution LST data with spatial integrity are particularly important for detecting the fluxes (such as soil heat flux) and surface properties (such as thermal inertia and urban heat island intensity) that are important to climate change, the urban thermal environment, and heat-related epidemiological studies (Fu et al. 2019; Liu and Weng 2012; Weng et al. 2014). However, how to obtain such LST data is facing a great challenge.

Thermal infrared (TIR) remote sensing represents a unique way to provide accurate long-term LST estimates over vast regions (Wu et al. 2019). However, TIR algorithms are severely constrained by cloud contamination and poor atmospheric conditions. For instance, about 60% of the MODIS LST products are cloud-contaminated, with higher percentages occurring at the low-mid latitudes (Cornette and Shanks 1993). Moreover, there is a tradeoff between the spatial and temporal resolutions in TIR remote sensing data, due to the hardware technology limitations and budget constraints (Zhan et al. 2013). For example, geostationary satellite sensors (e.g., MSG SEVIRI: ~ 3 km × 3 km and 15-min resolutions) have a high temporal resolution but a coarser spatial resolution, while polar-orbiting satellite sensors (e.g., Landsat Enhanced Thematic Mapper Plus (ETM+): 60 m × 60 m and 16-day resolutions) possess high spatial variability but loses temporal variability. The deficiencies in the TIR algorithms result in difficulties in obtaining spatially complete and high-spatio-temporal resolution LST data for use in potential applications.

With regard to the spatial incompleteness of LST data due to cloud contamination, a number of studies have aimed to fill the gaps in LST data (Shen et al. 2015). These gap-filling methods can be grouped into the following categories: 1) reconstruction methods dependent on spatial, temporal, or spatio-temporal information (Li et al. 2021; Pede and Mountrakis 2018; Wang et al. 2019; Yang et al. 2019; Zhao and Duan 2020); 2) surface energy balance (SEB)-based methods (Jia et al. 2021; Lu et al.

2011; Zeng et al. 2018); and 3) fusion-based methods (Duan et al. 2017; Long et al. 2020; Wang et al. 2014b). The first type of methods considers the neighboring LST pixels in space and time, and has been widely used for LST gap filling. Nevertheless, the reconstruction is usually based on clear-sky pixels and does not take the cloud cooling effect into account, thus resulting in a hypothetical clear-sky LST rather than real cloudy-sky LST (Zeng et al. 2018). In response to this issue, SEB-based methods have been proposed to estimate the actual LST under cloudy-sky conditions by considering related physical parameters (such as radiation and atmospheric forcing products). However, these physical process data are sometimes difficult to obtain and with a low accuracy in remote and less-developed regions. Compared with the other methods, the fusion-based methods do not necessitate additional auxiliary data, and have been proven effective in resolving the missing information of LST data (Zhang et al. 2021). A feasible approach is fusing TIR LST and passive microwave (PMW) LST which is available under cloudy conditions. For example, Zhang et al. (2019) introduced a method to merge the LST retrievals from PMW and TIR remote sensing based on temporal component decomposition. In addition, Xu and Cheng (2021) reported good LST reconstruction results by employing a fusion strategy combining cumulative distribution function matching and multiresolution Kalman filtering. However, PMW LST data adversely suffer from swath gaps and the swath depth, which greatly hinders the availability of continuous LST data (Duan et al. 2017).

With regard to obtaining high-spatio-temporal resolution LST data, the techniques fall into three main categories: 1) spatial downscaling methods; 2) temporal interpolation methods; and 3) spatio-temporal data fusion methods. The spatial downscaling methods are aimed at downscaling LST data with a coarse spatial resolution but frequent records, such as geostationary satellite data, by linking the spatial indicators (e.g., the normalized difference vegetation index (NDVI), the leaf area index (LAI), or the digital elevation model (DEM)) via linear/non-linear functions (Hutengs and Vohland 2016; Keramitsoglou et al. 2013; Zakšek and Oštir 2012). However, nearly no study has been able to downscale geostationary satellite LST data to a spatial resolution of greater than 100 m, because of the large scale difference (Quan et al. 2018). The temporal interpolation methods focus on interpolating temporally sparse thermal observations, but with high spatial details, typically for polar-orbiting satellites. The existing temporal interpolation methods mainly involve in diurnal temperature cycle (DTC) modeling and SEB modeling (Huang et al. 2014; Zhan et al. 2016). Despite obtaining a diurnal LST cycle, the spatial resolution usually remains unchanged (e.g., 1 km for MODIS LST) after temporal interpolation, which greatly constrains its application. Spatio-temporal data fusion (STF) methods have received much attention for leveraging the advantages of different data sources (Belgiu and Stein 2019). The basic principle of STF is integrating temporal change information from low-resolution data (e.g., MODIS imagery) and spatial detail information from high-resolution data (e.g., Landsat imagery), to obtain synthesized data with a high spatio-temporal resolution. Among the STF methods, the spatial and temporal adaptive reflectance fusion model (STARFM) (Gao et al. 2006) is the most widely used STF method. On the basis, a series of STF models have been proposed to improve the fusion accuracy in heterogeneous areas (Cheng et al. 2017; Huang and Zhang 2014; Liu et al. 2019a; Song and Huang 2012; Wang and Atkinson 2018; Zhu et al. 2010; Zhu et al. 2016b). Although STF methods were originally proposed for use with reflectance data, they appear to hold great potential for mapping LST. Most STF methods were initially devised to fuse LST data from two sensors (Yin et al. 2020; Zhu et al. 2021). However, in the case of a large scale difference, fusion from two sensors can result in step discontinuities or excessive smoothness in LST spatial pattern reconstruction (Xu and Cheng 2021). To alleviate this problem, Wu et al. (2015) developed a spatio-temporal integrated temperature fusion model to neutralize the negative impacts of large scale gaps by introducing intermediate-resolution data into the fusion process.

As an alternative, land surface models (LSMs) can provide long-term LST estimates with complete spatial coverage and a high temporal resolution (e.g., an hourly time scale), but usually with a low spatial resolution (e.g., 0.0625° × 0.0625° for China Land Data Assimilation System (CLDAS) LST and 0.25° × 0.25° for Global Land Data Assimilation System (GLDAS) LST). In addition, due to the indirect retrieval (i.e., modeling in a prognostic way) of the LSTs, the LSM output can reflect real rather than hypothetical LSTs on cloudy days, which can be selected as a promising background field for data fusion. A few studies have attempted to blend TIR and LSM-based LST to derive spatio-temporally complete LSTs. For example, Siemann et al. (2016) creatively merged the High-Resolution Infrared Radiation Sounder (HIRS) LST and NCEP Climate Forecast System Reanalysis LST to generate all-weather LST estimates at a 0.5° × 0.5° spatial resolution by the use of a Bayesian postprocessing procedure. To generate spatially complete LST data at a moderate/high spatial scale, Long et al. (2020) put forward a fusion framework combining MODIS LST and CLDAS LST based on the enhanced

STARFM (ESTARFM) algorithm. In addition, Zhang et al. (2021) blended GLDAS/CLDAS reanalysis LST and MODIS LST based on a merging method by decomposing LST time series. On the basis, Abowarda et al. (2021) incorporated Landsat LST into a two-step fusion approach to obtain gapless LST with 30-m and daily spatiao-temporal resolutions for soil moisture downscaling. These studies made full use of the LSM output to map gapless and moderate/high spatial resolution LST. However, the advantage of LSM-simulated LST in high temporal resolution was not fully exploited. As a result, the fused LST is commonly at a daily time scale, which cannot fulfill the need for diurnal fine-resolution LSTs for use in practical applications.

To the best of our knowledge, few studies have been able to provide real LST estimates with concurrent spatial continuity and high spatial and temporal resolutions (e.g., tens of meters and hourly scales). In this context, a satellite-LSM integrated temperature fusion framework is proposed for resolving the tradeoff between spatial completeness and the spatial and temporal resolutions in satellite-observed and LSM-simulated LST, with the aim being to reconstruct real, 60-m, and half-hourly all-weather LST data for both urban- and natural-dominated areas. The generated LST data under all weather conditions will be beneficial for the quantification of diurnal UHI intensity and will enable monitoring of the vulnerability of humans to vector-borne diseases.

## 2. Study area and data

### 2.1. Study area

Two typical urban- and natural-dominated areas were selected as the study areas (Fig. 1). The urban-dominated area is located in the center of the Wuhan metropolis in central China, spanning from 30°25' N to 30°42' N in latitude and from 114°11' E to 114°30' E in longitude (Fig. 1a, hereafter termed Wuhan_sub). The land use in this area is dominated by impervious surfaces (33.6%), cropland (34.1%), and water (21.8%). Wuhan has a subtropical monsoon climate, with the annual precipitation ranging from 1150 to 1450 mm and the maximum temperature in summer reaching up to 314 K. As one of the hottest "stove cities" in China, Wuhan is suffering from summer heat stress and heat-related health issues (Shen et al. 2016). It is therefore of special significance to conduct fine-scale heat-related studies in Wuhan.

The natural-dominated area is located in the middle reaches of the Heihe River Basin (HRB) within 38°43'N–39°00'N and 100°16'E–100°37'E in northwest China (Fig. 1b, hereafter termed Heihe_sub).

The land-cover types in this area are mainly composed by cropland (55.5%), bare land (18.7%), and grassland (10.4%). The climate of the HRB is dry, with annual precipitation of ~150 mm and annual mean temperature of ~281 K. The three weather sites used in the study are the Daman site (100.372°E, 38.856°N), the Zhangye wetland site (100.446°E, 38.975°N), and the Heihe remote sensing site (100.476°E, 38.827°N), which are part of the Heihe Integrated Observatory Network (details in Section 2.2.3). The observatory network in the HRB provides us with an ideal experimental field for the validation of land surface parameters (Liu et al. 2018; Xu and Cheng 2021).

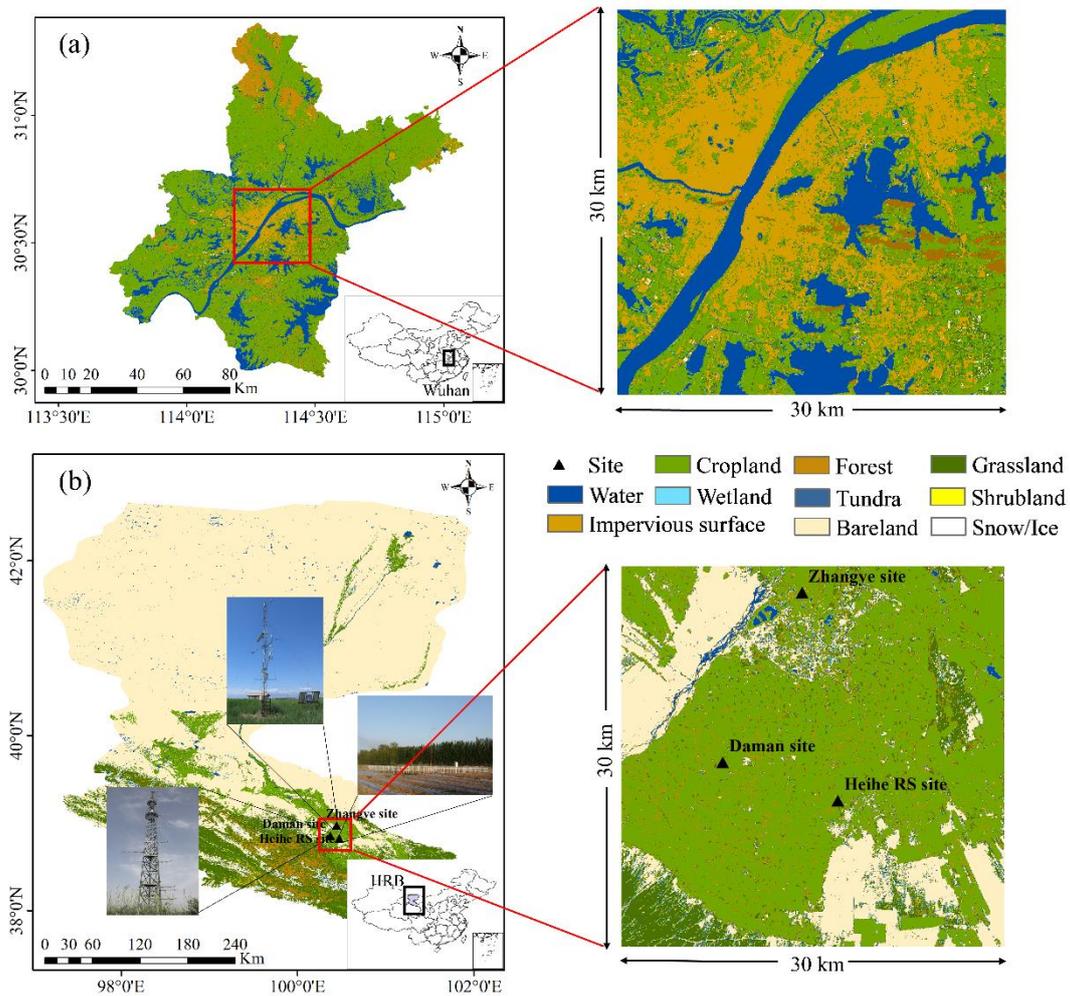

Fig. 1. Location of the two study areas (30 km × 30 km subsets), weather stations, and land-cover types in (a) Wuhan and (b) the HRB. The land-cover images with a 30-m resolution are from FROM_GLC_2015 (Wang et al. 2015).

2.2. *Data*

In this study, three kinds of data were used: 1) satellite data; 2) LSM simulations; and 3) in situ observations. Model-simulated LST and satellite-observed LST were used to generate LST data at a 60-m spatial resolution and a half-hourly temporal resolution. In situ LST data were used as the ground truth to validate the fusion results. Before fusion implementation, the model-simulated LST, MODIS LST, and Landsat LST needed to be resampled and re-projected to a common spatial grid and coordinate system. The details of the aforementioned data used in this study are provided in Table 1.

**Table 1**

The multi-source LST data from remote sensing, LSM simulations, and in situ measurements used in this study.

| LST data | Acquisition date | Acquisition time (UTC) |
| --- | --- | --- |
| Landsat ETM+ | 11 Oct 2013 | 02:52 |
| path: 123, row: 39 | 8 Aug 2013 | |
| MOD11A1 | 11 Oct 2013 | 03:14 |
| (h27v05) | 8 Aug 2013 | |
| Model-based LST | 8 Aug 2013 | Starting at 00:00 and every 30 min thereafter |
| Landsat ETM+ | 28 Feb 2016 | 03:58 |
| path: 133, row: 33 | 15 Mar 2016 | |
| MOD11A1 | 28 Feb 2016 | 04:12 |
| (h25v05) | 15 Mar 2016 | |
| | 13 Mar 2016 | 04:24 |
| | 17 Mar 2016 | 04:00 |
| Model-based LST | March 13, 2016–March 19, 2016 | Starting at 00:00 and every 30 min thereafter |
| Ground-based LST | March 13, 2016–March 19, 2016 | Starting at 00:00 and every 10 min thereafter |

*2.2.1. Satellite data*

We first collected the recently released Landsat-7 ETM+ Collection 2 Level-2 (L2C2) LST products covering the study area, which were acquired from the United States Geological Survey (https://earthexplorer.usgs.gov/). The Landsat-7 TIR band has a 60-m spatial resolution and a 16-day revisit cycle. Note that the Landsat LST data had been resampled to a 30-m resolution when downloaded, the fusion was therefore performed at this scale. The Landsat L2C2 LST product is generated from the Landsat Collection 2 Level-1 top of atmosphere (TOA) brightness temperature, TOA reflectance, the Advanced Spaceborne Thermal Emission and Reflection Radiometer Global Emissivity Dataset (ASTER GED), ASTER NDVI, and atmospheric profiles from reanalysis data using a single-channel algorithm (U.S. Geological Survey, 2021). To recover the missing pixels in the ETM+ scan line corrector-off (SLC-off) imagery, the weighted linear regression (WLR) based on a multi-temporal information method (Zeng et al. 2013) was applied in this study.

The MODIS Terra Land Surface Temperature/Emissivity Daily L3 Global 1 km SIN Grid

(MOD11A1) product in Collection 6 was collected. The MOD11A1 product provides LST estimates with an approximate view time of 10:30 a.m./10:30 p.m. (local solar time) in ascending/descending orbit, with a spatial resolution of 1 km × 1 km, and is obtained using the generalized split-window algorithm (Wan and Dozier 1996). The MOD11A1 v6 product has been found to have an accuracy of within 2 K in most cases (Wan 2014). Note that only MODIS LST data flagged as "good quality", according to the quality control (QC) flags, and with view angles less than 30°, were used in the LST fusion. The MYD21A1 emissivity product was used to calculate the broadband emissivity for the in situ LST estimation. Both of the MODIS products can be obtained from the NASA Earth Observing System Data and Information System (https://ladsweb.modaps.eosdis.nasa.gov/).

*2.2.2. Land surface model forcing and surface data*

The China Meteorological Forcing Dataset (CMFD) (He et al. 2020) was used to force the Community Land Model Version 5.0 (CLM 5.0). The CMFD was produced based on several meteorological forcing datasets and observations from 753 operational stations of the China Meteorological Administration (CMA) from 1979 to 2018, with a 0.1° × 0.1° spatial resolution and a 3-h temporal resolution. The dataset is composed of seven meteorological forcing variables: precipitation rate, air temperature, air pressure, specific humidity, wind speed, downward shortwave radiation and downward longwave radiation. The CMFD has been widely used in climate, hydrological, and diurnal land surface modeling studies, and is regarded as one of the best forcing datasets for China (Li et al. 2018). The surface data with a 0.05° × 0.05° resolution, such as percent plant functional types, percent urban, and percent lake, were acquired from the CLM surface data pool (Oleson et al. 2010).

*2.2.3. In situ data*

LST measurements from one superstation and two ordinary weather stations with different land-cover types in the Heihe Integrated Observatory Network were acquired to evaluate the accuracy of the fused LST. This integrated observatory network was established and is maintained by the Heihe Watershed Allied Telemetry Experimental Research (HiWATER) project. The Daman site (DM), Zhangye wetland site (ZY), and Heihe remote sensing site (HHRS) are equipped with four-component radiometers for providing surface upwelling and atmospheric downwelling longwave radiation every 10 min (available at http://data.tpdc.ac.cn/zh-hans/). Moreover, a homogeneity test for the three sites was conducted. We calculated the mean standard deviation (STD) of the Landsat LSTs at base and the predicted time of the fusion with subsets of 34 × 34 pixels within a MODIS LST pixel at the three weather stations. The results showed that the mean STDs of the LST at the DM, ZY, and HHRS sites were 0.94 K, 1.42 K, and 2.09 K, respectively, which reveals that the three sites possess reasonable spatial representativeness for a homogeneity validation. Table 2 lists the summary information for the three weather stations.

Ground-based LSTs were retrieved based on the method of Wang and Liang (2009), which can be expressed by the Stefan-Boltzmann law:

$$T_s = \left[\frac{L^\uparrow - (1-\varepsilon)L_\downarrow}{\varepsilon \cdot \sigma}\right]^{1/4} \quad (1)$$

where $T_s$ is the surface radiometric temperature or LST; $L^\uparrow$ and $L_\downarrow$ are the surface upwelling and atmospheric downwelling longwave radiation, respectively; and $\sigma$ is the Stefan-Boltzmann constant ($5.67 \times 10^{-8}$ W m$^{-2}$ K$^{-4}$). The broadband emissivity $\varepsilon$ was estimated from a linear regression equation (Wang et al. 2005) as follows:

$$\varepsilon = 0.2122 \cdot \varepsilon_{29} + 0.3859 \cdot \varepsilon_{31} + 0.4029 \cdot \varepsilon_{32} \quad (2)$$

where $\varepsilon_{29}$, $\varepsilon_{31}$, and $\varepsilon_{32}$ are the MODIS band 29, 31, and 32 narrowband emissivities.

**Table 2**

The details of the three weather stations used in this study.

| Site | Longitude | Latitude | Elevation (m) | Land cover | Instrument height (m) |
|---|---|---|---|---|---|
| DM | 100.372°E | 38.856°N | 1556 | cropland (maize) | 12 |
| ZY | 100.446°E | 38.975°N | 1460 | wetland | 6 |
| HHRS | 100.476°E | 38.827°N | 1560 | artificial pasture | 1.5 |

## 3. Methodology

Fig. 2 shows the overall framework of the developed approach, including two stages for generating LST estimates of spatial completeness at 60-m and half-hourly resolutions in two 30 km × 30 km areas. The first stage is the preprocessing before the data fusion, including CLM-based LST modeling and temporal and sensor normalization for the input LST data (see Section 3.2). The second stage is data fusion based on the preprocessed and normalized LST data using a filter-based spatio-temporal integrated fusion model, including the selection and weight calculation of similar pixels (see Section 3.3). Finally, the validation was conducted based on two evaluation schemes (see Section 4).

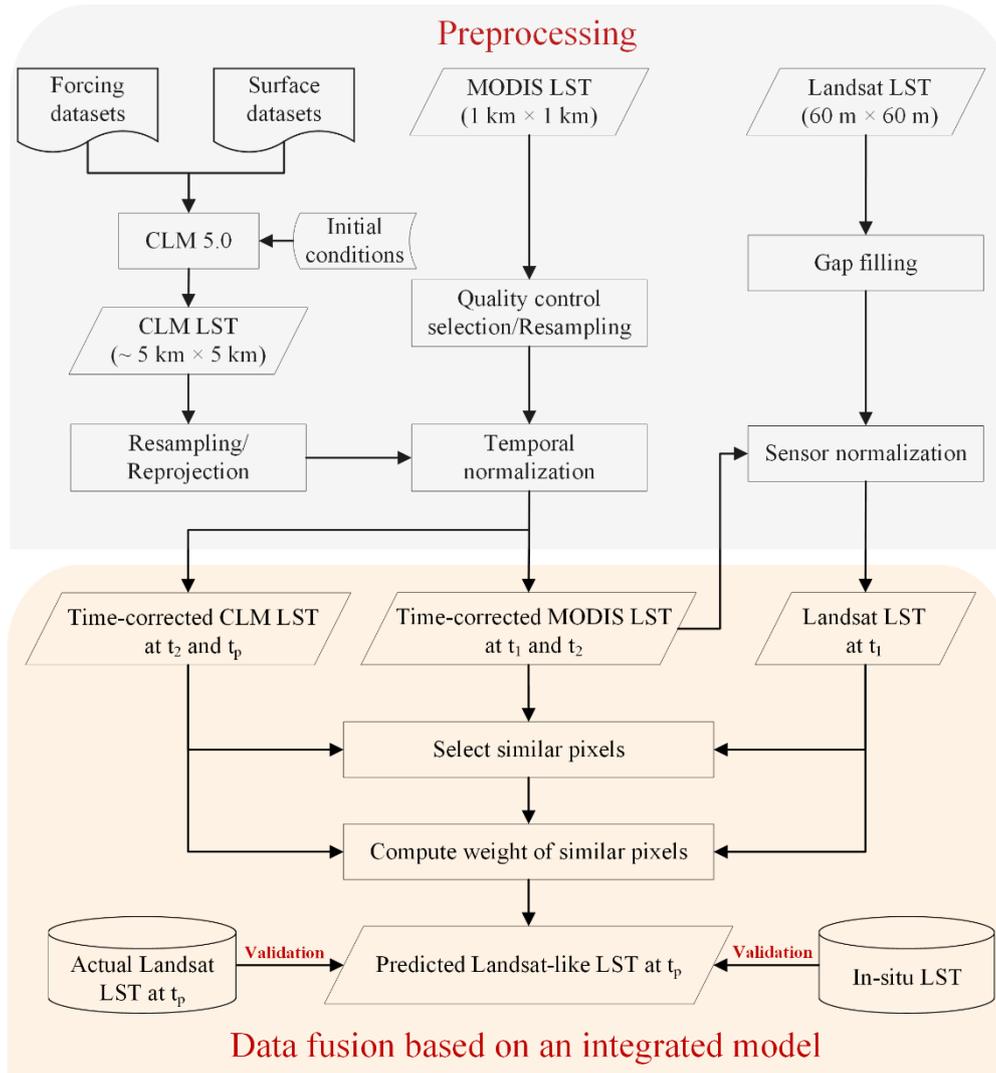

Fig. 2. The overall framework for generating gapless LST estimates with 60-m and half-hourly resolutions.

*3.1. High-resolution land surface modeling*

The Community Land Model Version 5.0 (CLM 5.0) is the latest global land model embedded in the Community Earth System Model (CESM2), which was developed by the CESM Land Model Working Group (LMWG) and the National Center for Atmospheric Research (NCAR) in 2019 (Lawrence et al. 2019). In CLM 5.0, spatial land surface heterogeneity is represented as a nested subgrid hierarchy in which grid cells are composed of multiple land units, snow/soil columns, and plant functional types (Lawrence et al. 2019). Compared with previous versions of CLM, CLM 5.0 contains notable improvements. The most important development is the introduction of a new soil evaporation resistance parameterization, which has been shown to strongly impact soil moisture and LST estimation (Chen et al. 2018; Deng et al. 2020; Swenson and Lawrence 2014). It worth noting that an urban parameterization scheme of CLM(CLMU) has been adopted since CLM 4.0, and a new building energy model has been integrated into the CLM 5.0 urban model, which can result in a better performance in LST simulation in urban areas (Oleson and Feddema 2020; Oleson et al. 2010). Serving as the state-of-the-art LSM, CLM 5.0 is becoming a new favorite in studies associated with land surface process modeling (Lombardozzi et al. 2020; Song et al. 2020).

In our study, CLM 5.0 was forced with the CMFD in an off-line mode. The default parameters in CLM 5.0 were adopted. Before the model simulation, a 10-year (2003–2012) spin-up experiment was run to obtain a stable initial condition for CLM 5.0. Simulations between January 1, 2013 and December 31, 2016 at a 30-min interval and with a spatial resolution of 0.05° × 0.05° (~ 5 km × 5 km) were then conducted for the two study domains.

*3.2. Temporal and sensor normalization between different LST data*

One prerequisite for STF is that the data pairs are comparable spatially and temporally, which calls for stability of the residual between data pairs at the reference time and target time (Gao et al. 2006). Specific to LST fusion, most of the residual results from two aspects, i.e., the view time difference and the sensor difference between the fine-resolution and coarse-resolution LST images. Hence, it makes great sense to conduct temporal and sensor normalization before LST fusion.

*3.2.1 Temporal normalization*

In our study, the view time of the Landsat-7 ETM+ (taking the Wuhan_sub study area as an example) is around 11:00 a.m. local solar time, while the view time of Terra MODIS daytime ranges from 10:00 a.m. to 11:42 a.m. local solar time. The existing mismatched view time between the MODIS LST and Landsat LST adversely hinders the implementation of data fusion. DTC models can be used to generate temporally continuous LST dynamics from discrete thermal observations, and are thus considered effective tools for temporal normalization (Duan et al. 2012; Zhu et al. 2016a). In this study, the GOT09 model, a semi-empirical DTC model combining extraterrestrial solar irradiation and the energy balance equation for the land surface, was utilized for the MODIS LST temporal normalization (Göttsche and Olesen 2009; Hong et al. 2018). The GOT09 model can be defined as follows:

$$\begin{cases} T_{day}(t) = T_0 + T_a \frac{\cos(\theta_z)}{\cos(\theta_{z,min})} \cdot e^{[m_{min}-m(\theta_z)]\tau}, t < t_s \\ T_{night}(t) = T_0 + \delta T + \left[T_a \frac{\cos(\theta_{zs})}{\cos(\theta_{z,min})} \cdot e^{[m_{min}-m(\theta_{zs})]\tau} - \delta T\right] e^{\frac{-12}{\pi k}(\theta-\theta_s)}, t \geq t_s \end{cases} \quad (3)$$

with

$$\theta = \frac{\pi}{12}(t - t_m) \tag{4}$$

where $T_{day}(t)$ and $T_{night}(t)$ are the daytime and nighttime LST at time $t$, respectively; $T_0$ is the residual temperature around sunrise; $T_a$ is the temperature amplitude; $\delta T$ is the temperature difference between $T_0$ and $T$; $t_s$ is the starting time of free attenuation; $t_m$ is the time when the temperature reaches its maximum; $\theta_z$ denotes the solar zenith angle; $\theta_{z,min}$ denotes the minimum zenith angle when $t = t_m$; $\theta_s$ denotes the thermal hour angle when $t = t_s$; $\theta_{zs}$ denotes the thermal zenith angle when $\theta = \theta_{zs}$; $m(\theta_{zs})$, $m_{min}$, and $m(\theta_z)$ are the relative air mass at $\theta_{zs}$, $\theta_{z,min}$, and $\theta_z$, respectively; $\tau$ is the total optical thickness (TOT); and $k$ is the attention rate of the nighttime temperature decrease. The formulas for calculating $\theta_z$, $m(\theta_{zs})$, and $k$ can be found in Göttsche and Olesen (2009).

The DTC model was initially developed based on sparse satellite thermal observations (e.g., MODIS LST). However, clouds are prevalent over most of the Earth's surface, which can lead to an inadequate number of MODIS LSTs (i.e., less than four) available for DTC modeling. A feasible solution is to use LSM-simulated LSTs instead to reconstruct the diurnal LST cycle, since the basic laws of the diurnal variation of LSTs are not limited to satellite-based LST (Huang et al. 2014; Jin and Dickinson 1999). In this study, CLM LSTs at base time ($t_1$) were used to generate the diurnal LST cycle (hereafter termed CLM_DTC) based on GOT09. The obtained CLM_DTC was then utilized to normalize the view time of the MODIS LST to be consistent with that of the Landsat LST, based on the assumption that the change of LST at the view times between MODIS and Landsat is approximately linear (Duan et al. 2014). The transformation can be defined as follows:

$$LST_M(t_{cor}, d) = LST_M(t_{ori}, d) + LST_c(t_{cor}, d)_{t_n} - LST_c(t_{ori}, d) \tag{5}$$

where $LST_M$ is the MODIS LST (resampled to 30 m); $LST_c$ is the CLM LST (resampled to 30 m) fitted by CLM_DTC; $t_{ori}$ and $t_{cor}$ are the view times before and after temporal normalization; and d denotes the date. At base time ($t_2$) and predicted time ($t_p$), the temporal linear interpolation method (i.e., interpolation between the two adjacent half-hourly CLM LST values) was used for the CLM LST temporal normalization. Fig. 3 shows the schematic of temporal normalization for the input LST in the case of Wuhan_sub.

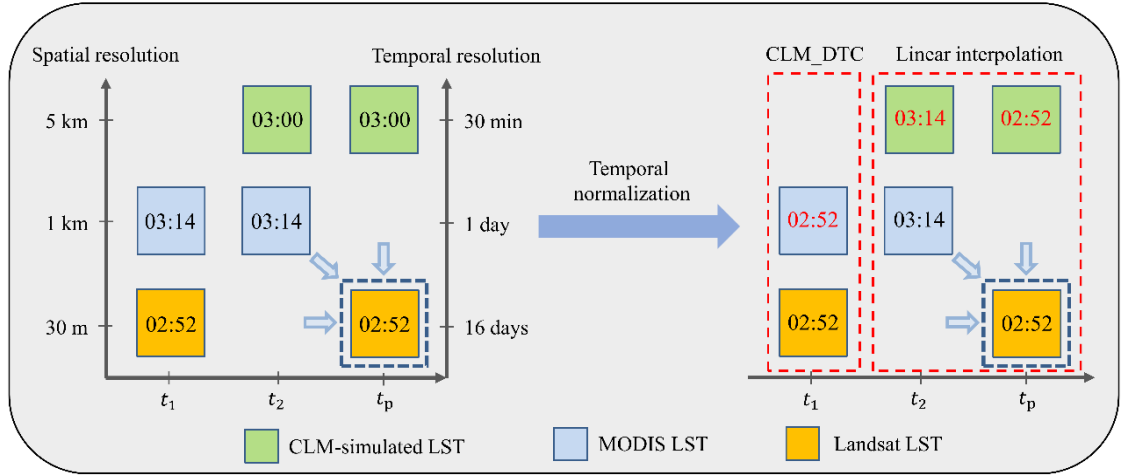

Fig. 3. Schematic of temporal normalization for MODIS and CLM LST in the case of Wuhan_sub. $t_1$ and $t_2$ denote the base times, while $t_p$ denotes the predicted time. The black and red numbers in the rectangles denote the view times in UTC before and after temporal normalization, respectively. Local solar time = UTC+7.6 h in Wuhan_sub.

*3.2.2 Sensor normalization*

Due to the different LST inversion methods, air conditions, viewing geometries, etc., the sensor bias between Landsat LST and MODIS LST is not constant. To eliminate or at least limit the bias between the sensors, the global linear model (GloLM)(Gan et al. 2014) was utilized for the sensor normalization in this study. Since the MODIS LST has been validated to be more reliable than the Landsat LST, we selected the MODIS LST as the reference axis (Shen et al. 2020). Note that the MODIS LST used here was the LST after temporal normalization based on the DTC model, so that it shared the same view time as the corresponding Landsat LST. It was assumed that there is a linear relationship between the MODIS LST and Landsat LST in the GloLM. The least-squares estimator was used to fit the parameters based on the point sets of pure pixels from the reference MODIS LST and aggregated Landsat LST (resampled to 1 km × 1 km). This linear relationship was then applied to the under-normalized Landsat LST pixels. To limit the scale effect in the process of sensor normalization, the purity of the coarse-resolution pixels was calculated based on the land-cover data.

*3.3. LST fusion with a filter-based spatio-temporal integrated fusion model*

In this study, we first attempted to blend the multi-source satellite-observed and model-simulated LST data through a filter-based spatio-temporal integrated fusion model, to generate gapless LST estimates with high spatial and temporal resolutions. The implementation of the integrated fusion model involves three main steps: 1) selection of similar pixels within a predetermined moving window; 2) determination of the spatio-temporal weighting function and calculation of the weights of similar pixels; and 3) estimation of the fine-resolution LSTs at the predicted time (Wu et al. 2015). On this basis, the integrated fusion algorithm was used to blend the Landsat LST map (60 m × 60 m), MODIS LST maps (1 km × 1 km), and CLM LST maps (~ 5 km × 5 km), to generate LST maps at 60 m × 60 m and half-hourly resolutions. The equations are given as follows:

$$L(x_{w/2}, y_{w/2}, t_p) = \sum_{i=1}^{N} W_i * \begin{pmatrix} L(x_i, y_i, t_1) - M(x_i, y_i, t_1) + \\ M(x_i, y_i, t_2) - C(x_i, y_i, t_2) + C(x_i, y_i, t_p) \end{pmatrix} \quad (6)$$

where $L$, $M$, and $C$ respectively represent the LST values from Landsat, MODIS, and CLM with successive descending resolutions; $w$ is the size of the sliding window; $L(x_{w/2}, y_{w/2}, t_p)$ is the final fused LST of the central pixel $(x_{w/2}, y_{w/2})$ within the sliding window at the predicted time of $t_p$; $(x_i, y_i)$ is the location of the $i$-th similar pixel; $N$ refers to the number of similar pixels; $t_1$ and $t_2$ are the base times; and $W_i$ refers to the spatio-temporal weighting function. It is noteworthy that only pixels similar to the central pixel in the fine resolution were selected to compute the LST. In this fusion algorithm, similar pixels are selected by the estimated number of land-cover classes and the standard deviation for the LST data, which can be expressed as:

$$|L(x_i, y_i, t_1) - L(x_{w/2}, y_{w/2}, t_1)| \leq \sigma * 2/m \quad (7)$$

where $L(x_i, y_i, t_1)$ and $L(x_{w/2}, y_{w/2}, t_1)$ respectively denote the neighboring pixel and the central pixel of the Landsat LST at the base time of $t_1$ within the sliding window; σ denotes the standard deviation for the Landsat LST; and $m$ denotes the estimated number of land-cover classes.

In the integrated fusion algorithm, the spatio-temporal weighting function $W_i$ consists of the similarity difference, the scale difference, and the geometric distance difference (Wu et al. 2015), which can be defined as follows:

$$W_i = \frac{1/(C_i \cdot SD_i)}{\sum 1/(C_i \cdot SD_i)} \quad (8)$$

$$SD_i = \frac{\exp(-|L(x_i, y_i, t_1) - L(x_{w/2}, y_{w/2}, t_1)|)}{\sum \exp(-|L(x_i, y_i, t_1) - L(x_{w/2}, y_{w/2}, t_1)|)} \quad (9)$$

where $SD_i$ is the degree of similarity between the central pixel and surrounding similar pixels in the sliding window of the fine-resolution data; and $C_i$ is the combination of the scale difference and the distance difference, which can be defined as follows:

$$C_i = \frac{\ln(R_i * 100 + 1) * D_i}{\sum \ln(R_i * 100 + 1) * D_i} \quad (10)$$

$$R_i = |L(x_i, y_i, t_1) - M(x_i, y_i, t_1) + M(x_i, y_i, t_2) - C(x_i, y_i, t_2) + C(x_i, y_i, t_p)| \quad (11)$$

$$D_i = 1 + \sqrt{(x_i - x_{w/2})^2 + (y_i - y_{w/2})^2} / (w/2) \tag{12}$$

where $R_i$ is the scale difference between the three different-resolution LST data sources. $D_i$ is the relative geometric distance between the $i\text{-}th$ similar pixel and the central predicted pixel.

## 4 Results

On the premise of preprocessing (i.e., temporal and sensor normalization) of the input data, two fusion scenarios were considered: in the first scenario, we fused the LST data from Landsat, MODIS, and CLM (hereafter termed L-M-C fusion), while in the second scenario, we fused only the Landsat and CLM LST (hereafter termed L-C fusion). Furthermore, to investigate the applicability of the developed approach for urban- and natural-dominated landscapes, two experiments (EXP1 and EXP2) were conducted. For EXP1 (Wuhan_sub), the fused LST values were evaluated only with the real Landsat LST at the predicted time, since there are no LST observations available for this area. For EXP2 (Heihe_sub), the fused LST values were jointly evaluated against real Landsat LST as well as in situ LST observations. For a quantitative evaluation of the LST predictions, four common statistical metrics are used in this paper: the Pearson correlation coefficient (R), the mean absolute error (MAE), the root-mean-square error (RMSE), and the overall bias (BIAS).

*4.2 Accuracy evaluation of the land surface modeling LST*

Before the data fusion, the CLM-simulated LST was first assessed using the MODIS LST. Fig. 3 shows a comparison between the CLM LST (resampled to 1 km) and the cloud-free MOD11A1 LST in the 2013 Wuhan_sub and 2016 Heihe_sub areas, for both daytime and nighttime. Compared with the MOD11A1 product, the CLM LST shows a reasonable accuracy, with R ranging from 0.94 to 0.98, MAE ranging from 1.59 K to 4 K, and BIAS ranging from −1.2 K to 1.54 K. It is also apparent that the accuracy of CLM LST at nighttime is better than that at daytime in regard to R and MAE, which is consistent with the conclusions of previous studies (Ma et al. 2021; Siemann et al. 2016). The above results indicate that the CLM 5.0 model has the potential to generate a consistent and reliable LST dataset, which could be chosen as a qualified background for the subsequent data fusion.

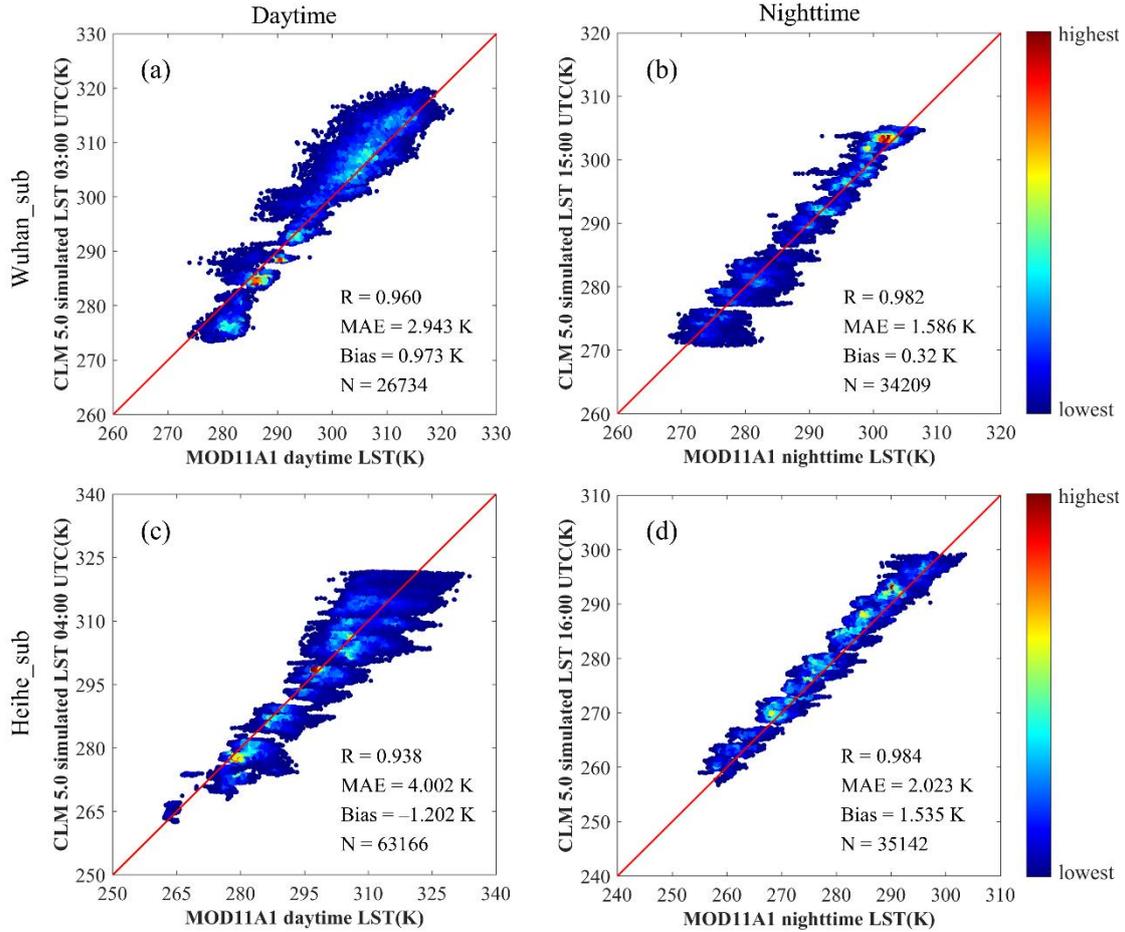

Fig. 4. Scatter plots between the MOD11A1 clear-sky LST (mean view time of ~10:30 a.m. and ~22:30 a.m. local solar time in the daytime and nighttime, respectively) and the corresponding CLM LST in the Wuhan_sub area in 2013 (a)–(b) and the Heihe_sub area in 2016 (c)–(d). The color bar denotes the density of the samples. N denotes the sample size. Local solar time = UTC+7.6 (6.7) h in Wuhan_sub (Heihe_sub).

*4.3 Experimental results for the urban-dominated area*

To investigate the feasibility of the proposed integrated fusion framework in a highly heterogeneous case, this experiment (EXP1) was aimed at deriving half-hourly "Landsat-like" LSTs in the urban-dominated area. Fig. 5 shows the observed and predicted LSTs for the data fusion in the Wuhan_sub area. Figs. 5a–c and e were used as the reference data for the L-M-C fusion, while the 48 (every 30 min) CLM LSTs on August 8, 2013 were used as the input data for deriving the half-hourly LST estimations. For simplicity, we only present the CLM LST at 02:52 UTC on August 8, 2013 in Fig. 5f. Note that the proposed fusion framework also allows for the fusion of two data sources. Fig. 5d was used as the coarse-resolution reference data to obtain the "Landsat-like" prediction in the L-C fusion. Figs. 5g and h show the L-C and L-M-C predicted LSTs, respectively. The observed ETM+ LST at 02:52 UTC on August 8, 2013 (Fig. 5i) was selected as the actual high-resolution LST for the validation.

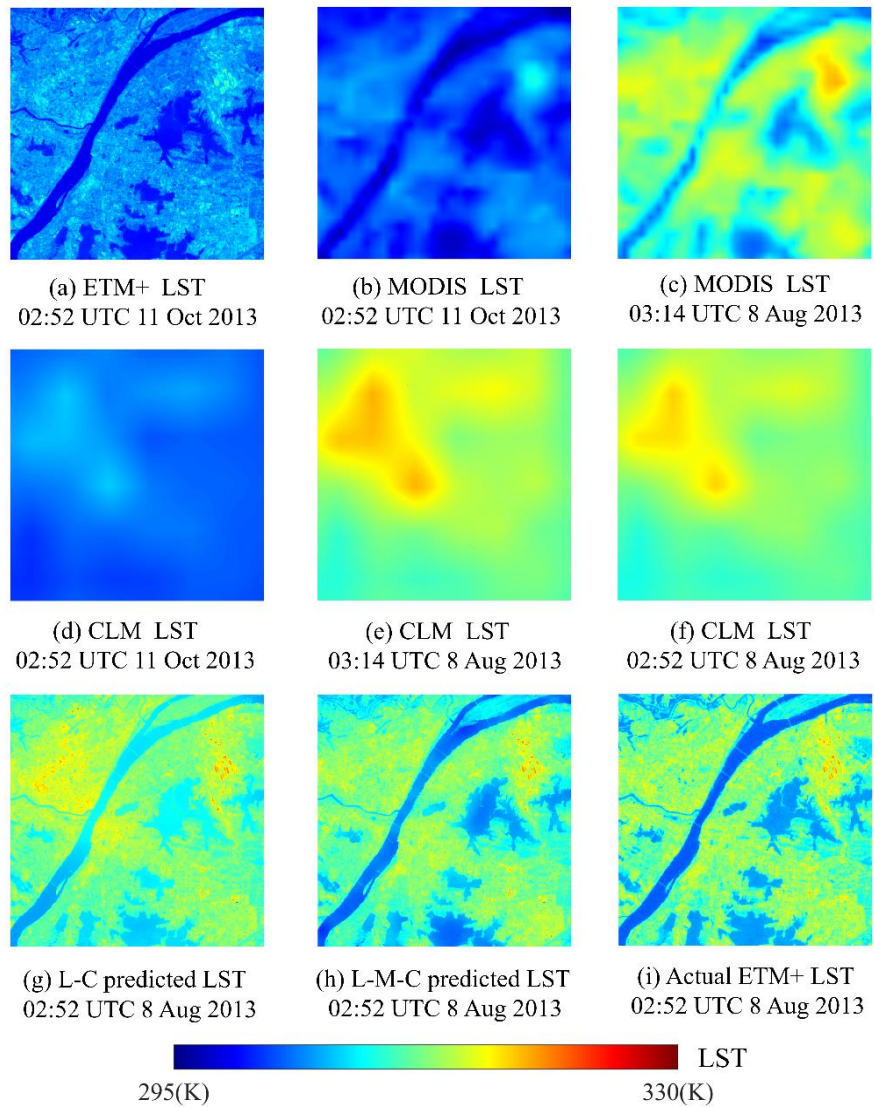

Fig. 5. Observed and predicted LSTs for the LST fusion in EXP1: (a) observed ETM+ LST on October 11, 2013; (b) observed MODIS LST after temporal normalization on October 11, 2013; (c) observed MODIS LST on August 8, 2013; (d) observed CLM LST after temporal normalization on October 11, 2013; (e) and (f) are the observed CLM LSTs after temporal normalization at 03:14 UTC and 02:52 UTC on August 8, 2013, respectively; (g) and (h) are the predicted "Landsat-like" LSTs under the L-C and L-M-C fusion scenarios on August 8, 2013; (i) observed actual ETM+ LST on August 8, 2013.

Overall, the two predicted LSTs resemble the actual Landsat LST reasonably well in regard to spatial patterns, but still showing a difference in LST magnitude. For a detailed visual judgment, two subsets of the actual and predicted LSTs under the two scenarios from Figs. 5g–i are shown in Fig. 6. When comparing the subsets, obvious overestimation can be observed in the L-C predicted LST, particularly over water and impervious surfaces. After integration of the MODIS LST in the L-M-C fusion, the predicted LSTs share considerable similarities with the actual LST in terms of spatial heterogeneity and LST magnitude, which largely makes up for the LST overestimation.

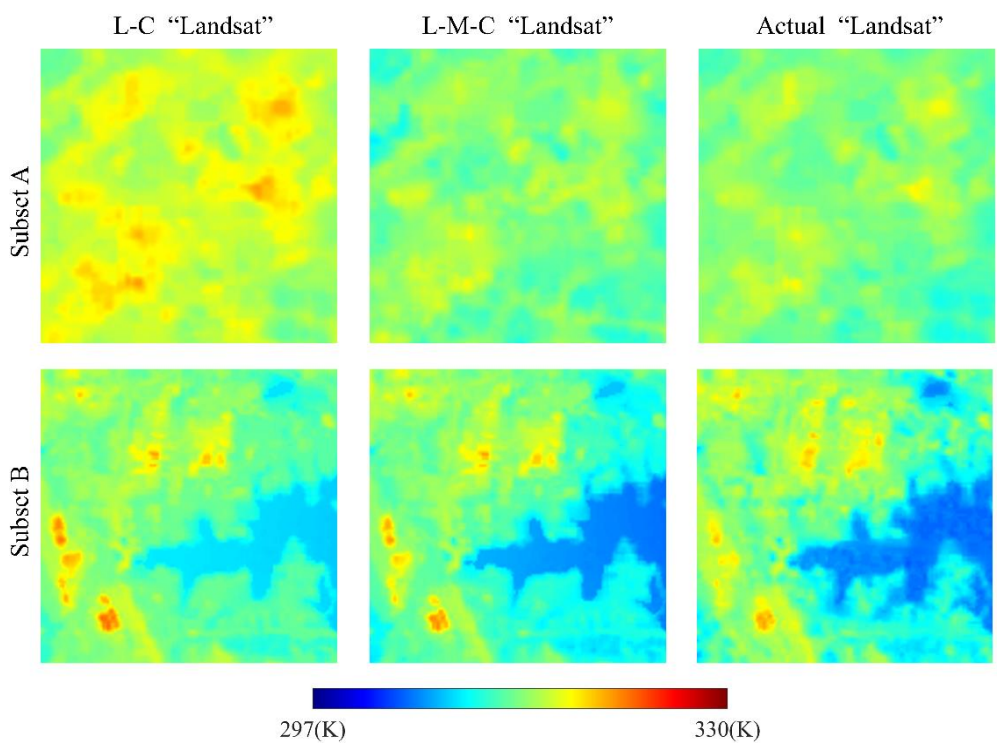

Fig. 6. Detailed comparison between the observed LSTs and predicted LSTs from Figs. 5g–i. The columns from left to right denote subsets of the L-C, L-M-C predicted LSTs and the actual LSTs, respectively. The first and second rows denote subsets A and B in the northwest and eastern middle of the Wuhan_sub area.

For a quantitative assessment, Fig. 7 shows the scatter density plots of the actual Landsat LST and predicted Landsat-like LST under the two scenarios on August 8, 2013. General agreement can be observed in the L-M-C prediction, with an R of 0.94, RMSE of 1.26 K, MAE of 0.98 K, and a small BIAS value, indicating the reliability of the proposed approach in Landsat-like LST prediction (Fig. 7b). Fig. 7 also shows the superiority of L-M-C fusion over L-C fusion across all the metrics. This phenomenon can be largely explained by the introduction of the moderate-resolution MODIS LST in the data fusion process, which can neutralize the negative impacts of the large-scale difference between the CLM LST and Landsat LST.

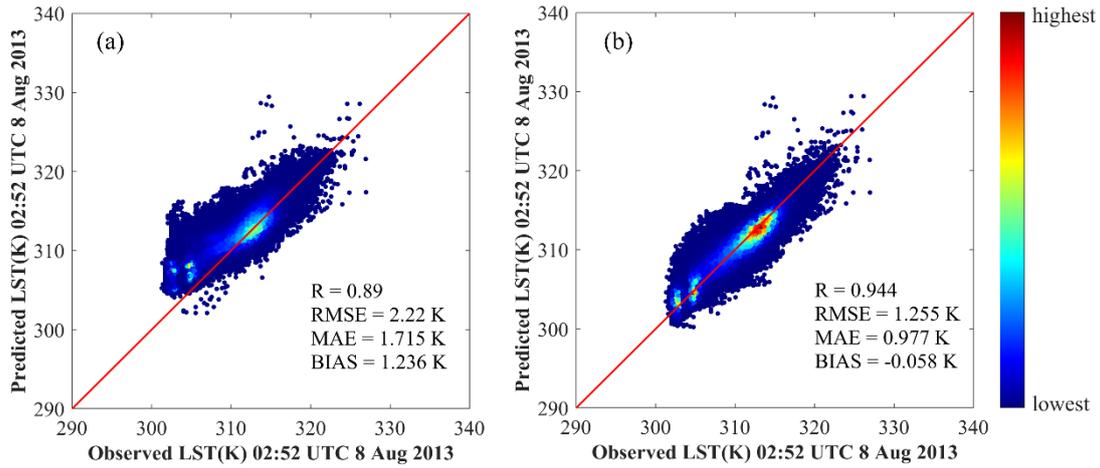

Fig. 7. Scatterplots of actual Landsat LST versus (a) the L-C predicted LST and (b) the L-M-C predicted LST in EXP1 on August 8, 2013. The color bar denotes the density of the samples.

To better depict the LST diurnal cycle, Fig. 8 displays the 48 predicted LSTs based on L-M-C fusion for August 8, 2013. Benefiting from the spatial continuity of the background data (i.e., the CLM LST), the fusion results successfully inherit gapless features. Moreover, the predicted diurnal LSTs contain the vast majority of the spatial details found in the observed images, including surface features such as rivers and lakes with low LSTs and urban areas with high LSTs. Notable LST variation over the course of the day can also be observed. The diurnal LST peak is observed at UTC 06:00 (local time 13:42), which is in accordance with the results of other studies (Gao et al. 2019). Overall, the proposed integrated temperature fusion framework is able to reproduce diurnal LSTs in magnitude and time evolution reasonably well for landscapes with high heterogeneity.

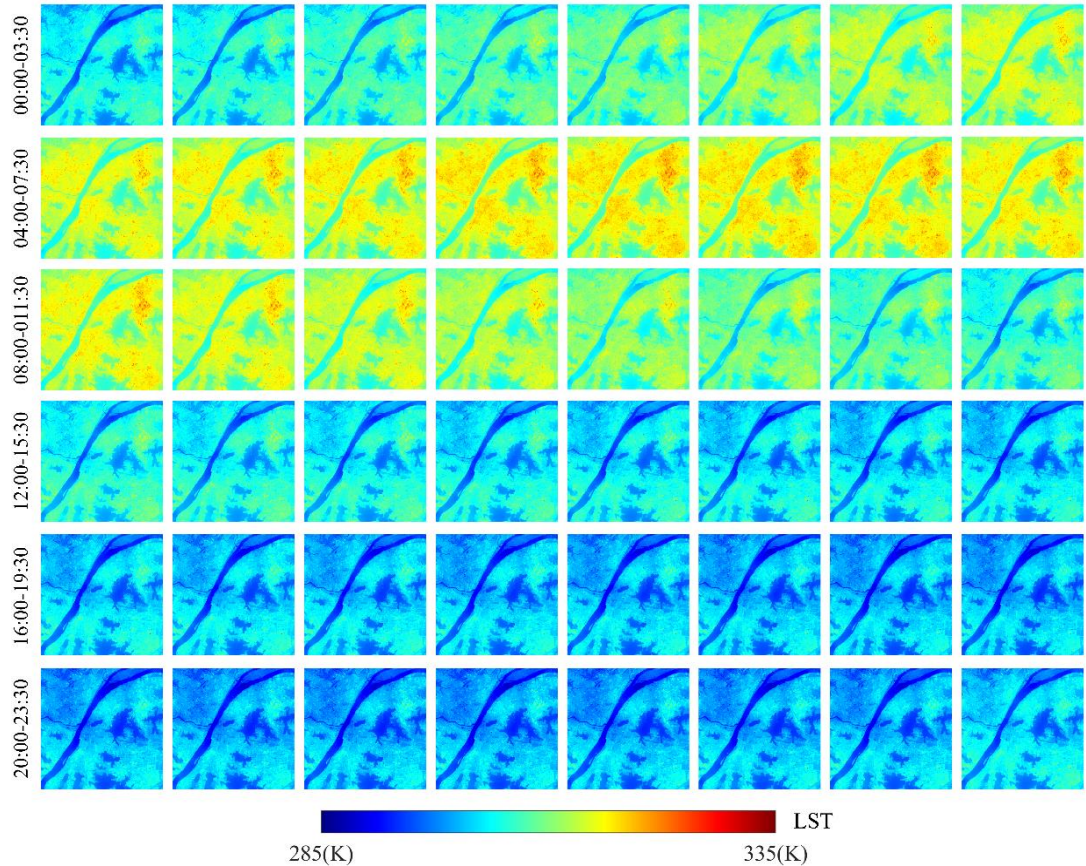

Fig. 8. The 48 predicted gapless Landsat-like LSTs at a half-hourly temporal resolution for the Wuhan_sub area on August 8, 2013.

*4.4 Experimental results for the natural-dominated area*

In this experiment (EXP2), the integrated fusion framework was tested over a natural-dominated area. The observed and predicted LSTs for the data fusion in EXP2 (Heihe_sub) are shown in Fig. 9. Similarly, for the L-M-C fusion, the data from Figs. 9a–c, and e served as the input base data. For the L-C fusion, only Figs. 9a and d were selected as the input base data, while Fig. 9f was used as the input coarse-resolution data at the predicted time for both fusion scenarios. The corresponding L-M-C and L-C predictions at the predicted time are shown in Figs. 9g and h, respectively. The 48 "Landsat-like" LST predictions for March 15, 2016 with a 30-min time step were also obtained in a similar manner, but are not shown here due to the space limitations. Fig. 9i was chosen as the actual LST for the validation.

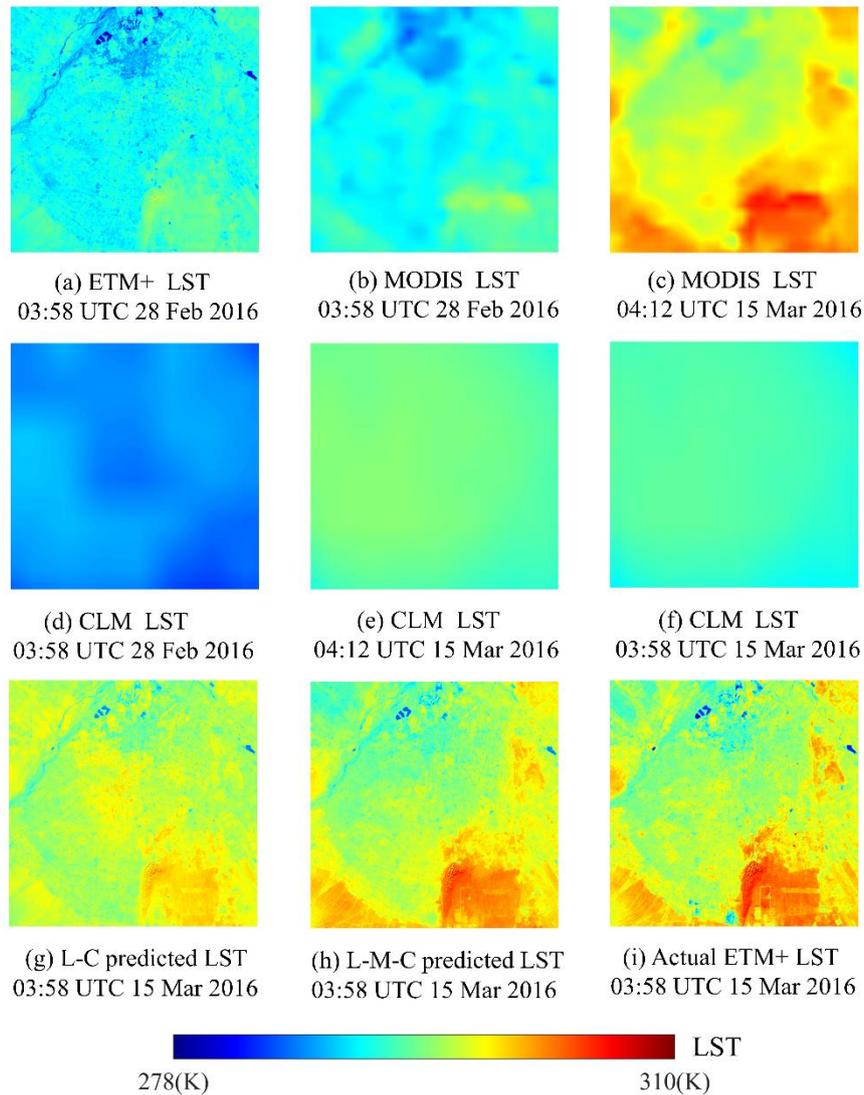

Fig. 9. Observed and predicted LSTs for LST fusion in EXP2: (a) observed ETM+ LST on February 28, 2016; (b) observed MODIS LST after temporal normalization on February 28, 2016; (c) observed MODIS LST on March 15, 2016; (d) observed CLM LST after temporal normalization on February 28, 2016; (e) and (f) are the observed CLM LSTs after temporal normalization at 04:12 UTC and 03:58 UTC on March 15, 2016, respectively; (g) and (h) are the predicted "Landsat-like" LSTs under L-C and L-M-C fusion scenarios on March 15, 2016; (i) observed actual ETM+ LST on March 15, 2016.

It is noticeable that the L-M-C prediction is considerably more in agreement with the actual observations in terms of LST magnitude and spatial details, while the L-C prediction shows an overall underestimation, especially on the bare land class (Figs. 9g–i). Error histograms of the LST predictions versus the actual Landsat LST under the two scenarios are shown in Fig. 10. The error distribution of the L-M-C fusion is much better than that of the L-C fusion, with mean values for the L-M-C fusion and L-C fusion of –0.13 K and –0.41 K and STDs of 0.96 K and 1.85 K, respectively. The MODIS LST was not only used to improve the LST details that cannot be captured by the CLM LST, but was also used as a complement to provide the temporal variation and constrain the fusion process, ultimately improving the LST fusion.

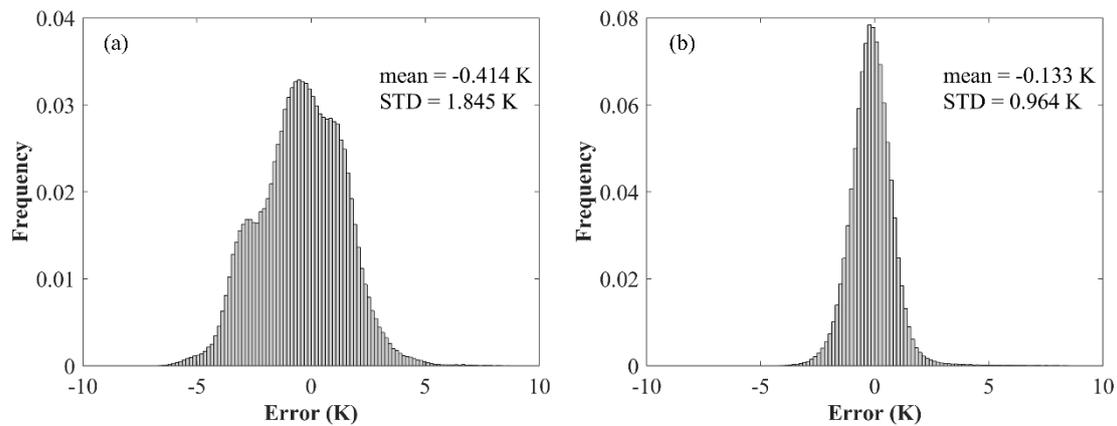

Fig. 10. Error histograms of (a) the L-C predicted LST and (b) the L-M-C predicted LST versus actual Landsat LST in EXP2 on March 15, 2016.

To evaluate the predicted LSTs in the diurnal cycle, ground-truth LSTs from three sites with different land-cover types were used to validate the performance of the proposed method. Fig. 11 shows a comparison of the L-M-C predicted LSTs and L-C predicted LSTs with the in situ LSTs. It can be seen that overall agreement can be observed for all the predictions at the three sites in terms of R greater than 0.98, indicating the feasibility of diurnal LST cycle modeling using LSM-simulated LSTs as the background field. Furthermore, the L-M-C predicted LSTs are closer to the in situ LSTs than the L-C predicted LSTs, especially at the DM and ZY sites. The prediction error of the L-M-C fusion (MAE = 0.82–2.6 K) is lower than that of the L-C fusion (MAE = 2.58–3.2 K), reflecting again the necessity of integrating moderate-resolution LST to serve the role of "scale transition" in data fusion. The underestimation of the LST peaks at the HHRS site can be mainly attributed to the relatively large-scale difference between the ground-measured LST and Landsat LST (Fig. 11c), which has also been observed in other studies (Wang et al. 2021a). Overall, the above results demonstrate that it is viable to obtain diurnal LSTs with a high spatial resolution based on the proposed integrated temperature fusion framework.

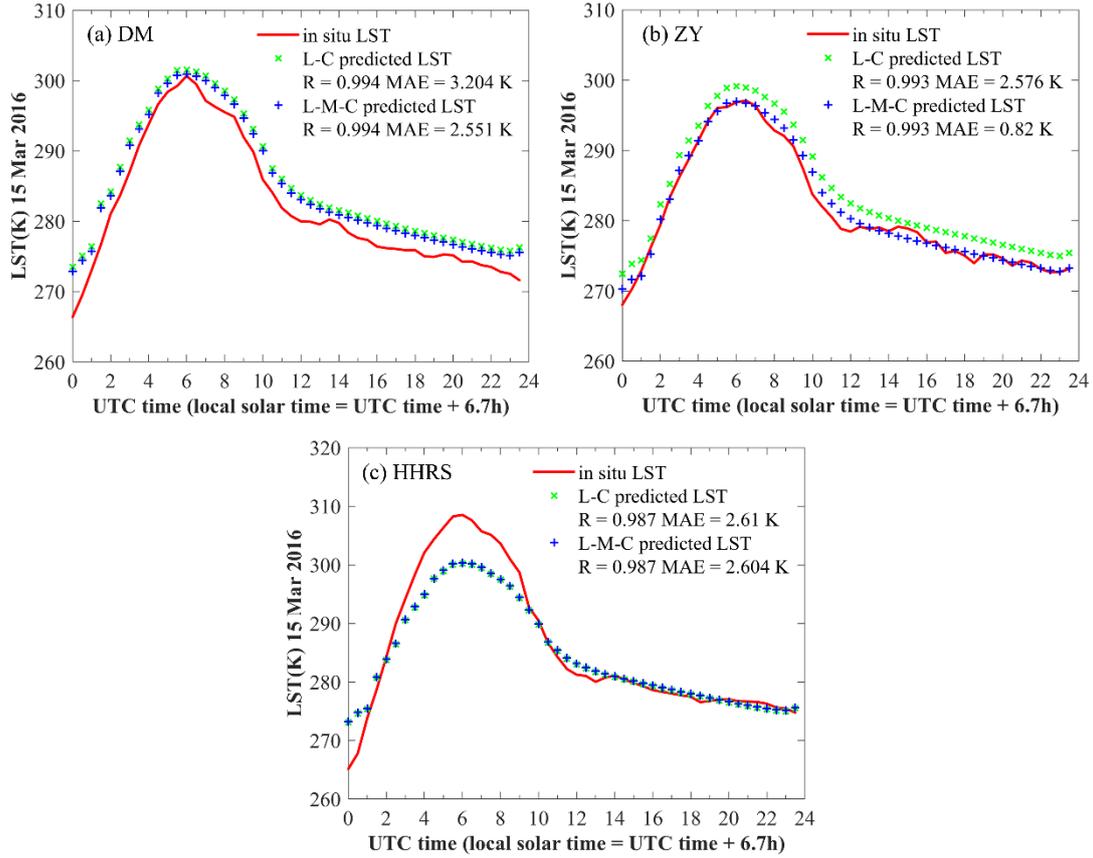

Fig. 11. Diurnal cycle of LSTs obtained from the L-C (green crosses) and L-M-C (blue pluses) fusion scenarios compared with that of the in situ LSTs (red line) on March 15, 2016 at the three sites, i.e. (a) the Daman site, (b) the Zhangye site, and (c) the Heihe remote sensing site.

To further analyze the transferability of the proposed approach for other predicted (consecutive) dates, Fig. 12 shows the time-series variation of the L-M-C and L-C predicted LSTs versus the ground-measured LSTs at the three weather sites during March 13, 2016–March 19, 2016. The high-quality (i.e., cloud free and within 30° view angles) MODIS LST is only available on March 13th, March 15th, and March 17th. The diurnal LST estimates on these dates were produced based on L-M-C fusion using separate MODIS LST as the reference data (i.e., reference moderate-resolution LST data at $t_2$). Meanwhile, on the other dates, the MODIS LST acquired on an adjacent date was used to generate the diurnal LST predictions, even though this does introduce some additional error. As a result, we produced the consecutive diurnal LST estimates for March 13, 2016–March 19, 2016.

From Fig. 12, it can be found that the L-M-C predicted LST at the three sites is in good consistency with the in situ LST measurements, and the diurnal LST variation at each site is basically reproduced well. Table 3 further lists the statistical metrics for the comparison results. For the L-M-C predictions, the MAE (RMSE) varies from 1.86 K (2.39 K) to 3.17 K (3.97 K), the BIAS varies from 0.09 K to 2.41 K, and R ranges from 0.97 to 0.98, indicating the robust performance of the proposed method in LST time-series reconstruction. In general, better agreements can be observed on the dates on which the available MODIS LST was involved in the fusion, demonstrating the important role of MODIS LST in the integrated fusion process. However, the LST reconstruction results obtained by L-M-C fusion do not match well with the in situ LST at certain times, especially for the lowest temperatures at the DM and HHRS sites (such as around 00:00 UTC, DOY: 77–78) (Figs. 12a and c). It is demonstrated that CLM model tends to overestimate LST at night due to the deficiency of the parameterization scheme, which is also shown in Fig. 4d, and the overestimation can be transmitted to the subsequent fusion process. Wang et al. (2014a) and Trigo et al. (2015) also reported obvious warm nighttime bias over most global land areas in LSM-based LST.

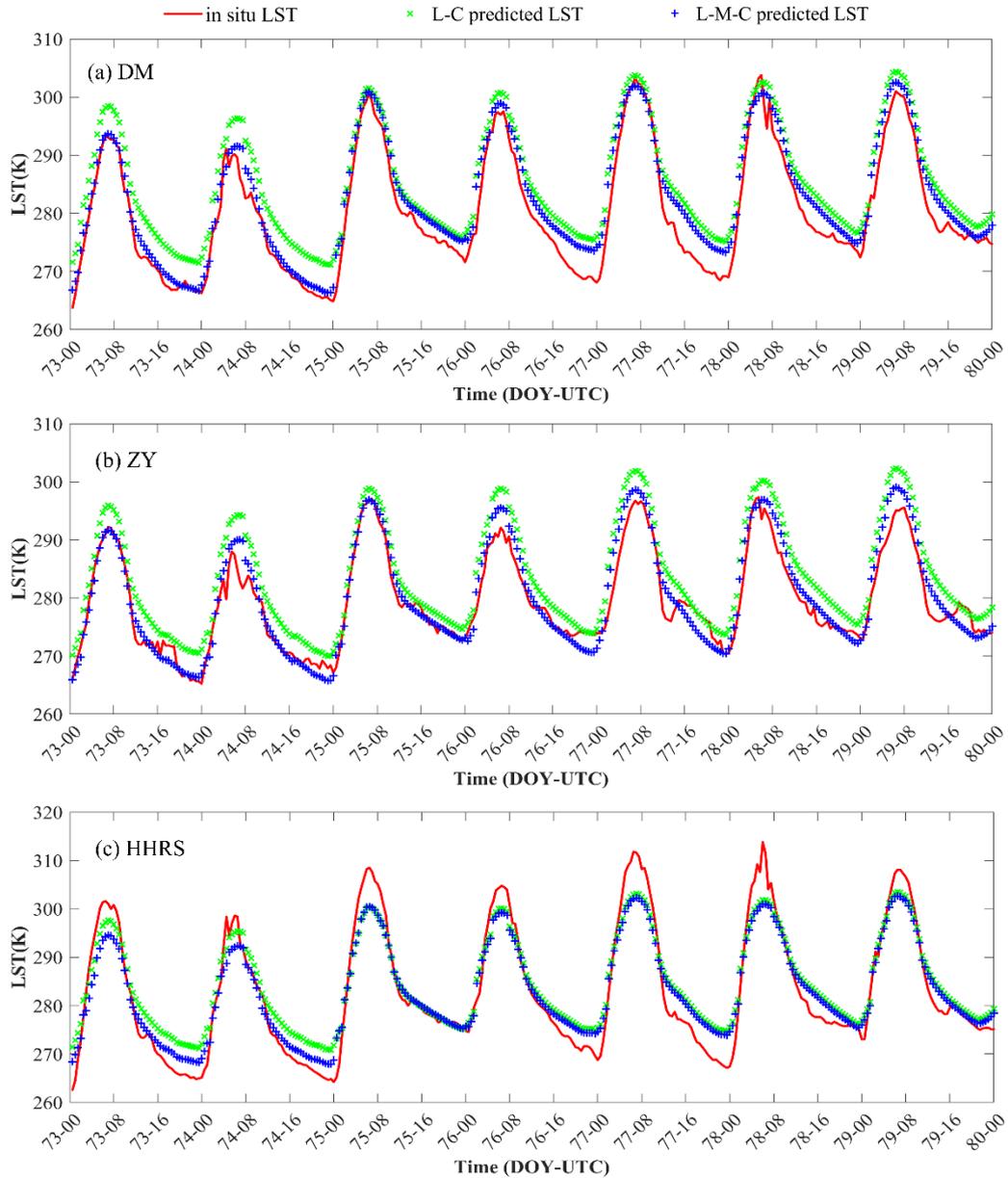

Fig. 12. Comparison between the L-C and L-M-C predicted diurnal LST time series versus the in situ LST measurements at the three sites, i.e. (a) the Daman site, (b) the Zhangye site, and (c) the Heihe remote sensing site during March 13, 2016–March 19, 2016. DOY denotes the day of year.

Compared with the L-M-C fusion, the L-C fusion demonstrates a relatively poor ability in capturing the LST peaks of the diurnal cycle and shows severe overestimation of the LST, especially for the DM and ZY sites, resulting in a poor performance across all the metrics (Table 3). The performance of the spatio-temporal LST fusion was found to be directly associated with the site domain characteristics, while the best and worst performances were respectively observed at the ZY site in the wetland area and the HHRS site in the artificial pasture area.

**Table 3**

Comparison of the metrics between the L-C and L-M-C predicted LSTs versus in situ LST measurements at the three sites during March 13, 2016–March 19, 2016.

| Site | R | | RMSE (K) | | MAE (K) | | BIAS (K) | |
|---|---|---|---|---|---|---|---|---|
| | L-C | L-M-C | L-C | L-M-C | L-C | L-M-C | L-C | L-M-C |
| DM | 0.983 | 0.983 | 5.315 | 3.061 | 4.956 | 2.583 | 4.94 | 2.413 |
| ZY | 0.97 | 0.974 | 4.674 | 2.391 | 4.057 | 1.861 | 4.035 | 0.667 |
| HHRS | 0.981 | 0.981 | 4.394 | 3.967 | 3.764 | 3.173 | 1.422 | 0.089 |

## 5  Discussion

*5.1  Perspectives on the view time inconsistency between the different LST data*

In the usual study, the view time matching of the base data pairs in the spatio-temporal fusion will be only approximate (e.g., the view times are approximately 10:30 a.m. for MOD11A1 daytime and 11:00 a.m. local solar time for Landsat), which can introduce errors into the subsequent fusion process. In this study, we utilized a DTC model to perform the correction. To further evaluate the influence of the temporal normalization on the fusion accuracy, we conducted a comparison of fusion with temporal normalization (hereafter termed Tnor fusion) and without temporal normalization (hereafter termed No_Tnor fusion) versus actual Landsat LST in a Landsat-scale evaluation of the two regions. The results indicate a slight improvement with the Tnor fusion, with the RMSE decreasing from 0.98 K–1.27 K to 0.97–1.26 K. Furthermore, we conducted a comparison in the diurnal cycle. Fig. 13 shows the comparison of the Tnor and No_Tnor predicted diurnal LSTs versus the in situ LST measurements at the three sites. A large improvement is found with the Tnor prediction at the DM and ZY sites, in terms of the RMSE decreasing from 3.92 K (1.53 K) to 2.78 K (1.09 K) and the MAE decreasing from 3.77 K (1.22 K) to 2.55 K (0.82 K) at the DM (ZY) site. Comparable results are found in the Tnor and No_Tnor fusion at the HHRS site, followed by a decrease of MAE and an increase of RMSE, which is partly due to the relatively high heterogeneity found at the HHRS site. It can be explained that after the implementation of the temporal normalization, the bias due to the view time difference between the reference data pairs is largely eliminated, thus improving the fusion accuracy, but this improvement may occasionally be minimal (Liu et al. 2019b). The magnitude of the improvement relies on the gap size of the view time difference and the quality of the satellite-derived LSTs themselves.

The above results indicate that it is effective to perform temporal normalization to maintain, or even improve the fusion accuracy, and it definitely enhances the generalization ability of spatio-temporal fusion, especially for the case of a large view time gap. Nevertheless, we have to mention that it can also introduce some uncertainty, even if such uncertainty can be neutralized by the benefits it brings. CLM_DTC normalization is implemented based on the assumption that the diurnal LST cycle generated by the LSM-simulated LSTs is similar to that of the satellite-observed LSTs. However, there are differences between the data sources due to the different LST retrieval mechanisms. Furthermore, the uncertainty can also be associated with the coarse resolution of the CLM-simulated LST, which may yield a relatively coarse LST difference between the times before and after the temporal normalization, and consequently a deviated prediction when the difference is added back to the MODIS LST to be normalized (Zhao et al. 2019).

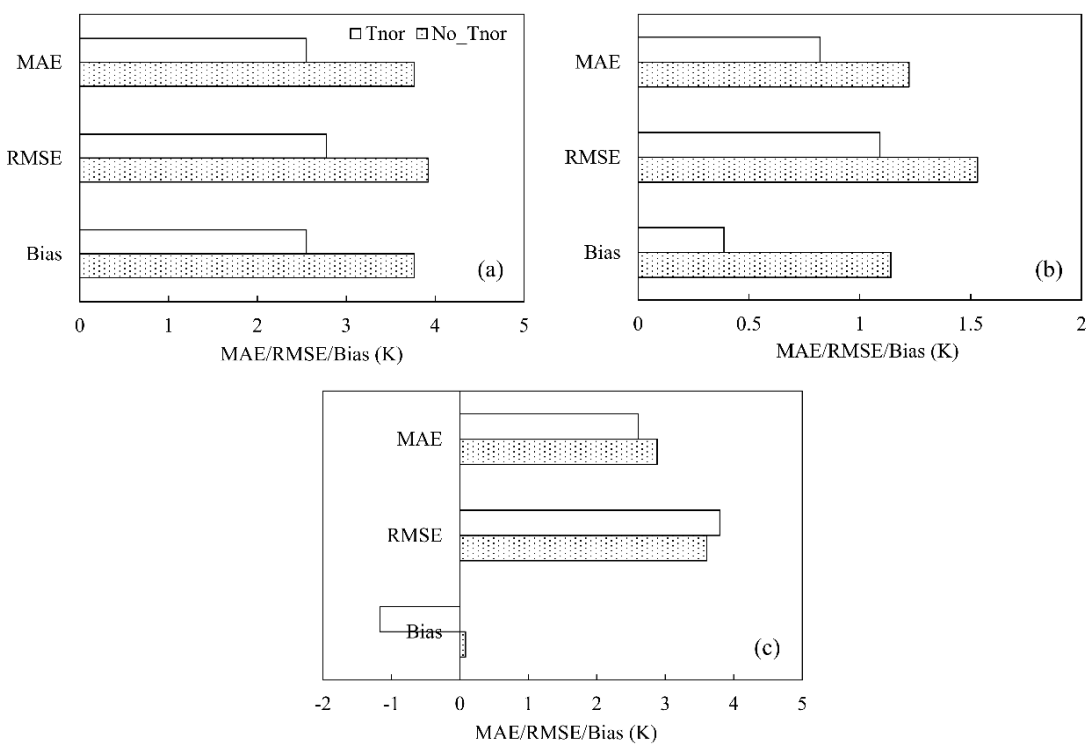

Fig. 13. Comparison of the Tnor and No_Tnor predicted diurnal LST cycle versus the in situ diurnal LST cycle at the three weather sites, i.e. (a) the Daman site, (b) the Zhangye site, and (c) the Heihe remote sensing site on March 15, 2016 in EXP2.

*5.2 Perspectives on the LSM simulation uncertainty*

As a background, LSM-simulated LST plays an important role in providing temporal change information for data fusion and temporal normalization. The accuracy of the LSM-simulated LST is highly dependent upon the parametric scheme of the model and the quality of the input atmospheric forcing dataset and surface dataset (Lawrence et al. 2019). The latest LSM (i.e., CLM 5.0) was applied to simulate real and high-quality LSTs in this study. However, as investigated in Ma et al. (2021), the performance of LST modeling in CLM is strongly related to the sensible heat roughness length scheme and soil thermal conductivity scheme, especially on bare ground surfaces, which needs to be further improved in CLM 5.0. The CMFD is regarded as one of the best atmospheric forcing datasets in China (He et al. 2020). However, the distribution of the operational CMA stations used in the CMFD is severely uneven, with few stations in western China (covering Heihe_sub). This increases the uncertainty of the CMFD over this region, consequently leading to deviations in the simulated LST. Uncertainties may also arise from the static and outdated land surface datasets used in CLM 5.0. In light of this issue, some scholars have attempted to apply bias correction to the LSM-based LST, to make them comparable to fine-resolution LST data (e.g., MODIS LST) (Long et al. 2020; Zhang et al. 2021). However, in this study, we focused on deriving half-hourly LSTs, where there were no corresponding diurnal high-accuracy LSTs as reference data for bias correction. Future studies can be expected to conduct such bias correction by introducing geostationary satellite data.

*5.3 Perspectives on the MODIS view angles*

In this study, the experiments were conducted only for a single day (EXP1) and a week (EXP2). In theory, the proposed method could also be used to produce more consistent LST estimates (e.g., one month or longer). However, in addition to the overpass time, MODIS LST is also sensitive to the view angle and sun geometry. As investigated in Wang et al. (2021b), the anisotropy over the MODIS view angles can reach up to 6.6 K for Terra day (around 11:00 a.m. local time) in urban areas. In the generation of such a time series of LST, it is necessary to address the angular anisotropy in the MODIS LST (Cao et al. 2019). Although some studies have attempted to conduct correction of the angular effects through semi-empirical or physical models, they have usually been subject to the requirement for accurate prior knowledge (e.g., on canopy structure or spectral properties) or sufficient input parameters, which badly hampers their practical application (Cao et al. 2021). To restrict the angular effect, only MODIS LST obtained under small viewing angles (less than 30°) was used in this study. Future studies will need to explore a more practical method for satellite-derived LST angular normalization, to further extend the fusion condition and produce longer time series of LST estimates.

*5.4 Perspectives on the land-cover change and scaling effects*

The proposed satellite-LSM integrated fusion model has been shown to be effective in generating gapless LSTs with high spatial and temporal resolutions. However, it has a limitation in predicting LST in the case of land-cover type change. An unmixing-based method could be incorporated into the integrated model for predicting land-cover changes. For example, Quan et al. (2018) proposed a linear temperature mixing model (LTMM) to account for the impacts of landscape heterogeneity and land-cover type change. Another solution is to add one or more spatially complete Landsat-MODIS LST pair into the integrated fusion process, as in the research into ESTARFM (Zhu et al. 2010), even though it could be hard to find such data pairs over a close time period in practical applications (Zhao et al. 2018).

The scaling effects caused by spatial mismatch can be a significant problem (Bei et al. 2020). Despite the relatively high spatial resolution (i.e., 5 km × 5 km) of CLM-simulated LST compared with other reanalysis LST data, the land surface heterogeneities can still be hardly represented, which may cause a discrepancy in CLM-simulated diurnal LST patterns. Although we have confirmed the spatial representativeness of the ground LST measurements in the corresponding 1-km LST for validation, the point-to-pixel LST match cannot be totally guaranteed. The four-component radiometer heights at the HHRS, ZY, and DM sites are 1.5 m, 6 m, and 12 m, respectively, which has a major impact on the footprint size. The weather sites can only represent a scale of several to tens of meters, while Landsat-like LST estimation is at a 60-m scale, thus resulting in a discrepancy in the comparison of fused LST and in situ LST.

## 6 Conclusion

To date, few studies have been able to provide real LST estimates with concurrent spatial completeness and high spatio-temporal resolutions. In this paper, we proposed a satellite-LSM integrated temperature fusion framework aiming at generating real, gapless, and high spatio-temporal resolution (i.e., half-hourly, 60-m resolution) LST data, which is not achievable through the use of a single technique. After temporal and sensor normalization, the fused LST estimates on predicted dates showed good agreement with the high-resolution Landsat LSTs over both urban- and natural-dominated regions, in terms of an R value of 0.94 and MAE of 0.71–0.98 K. Furthermore, the generated time series of all-weather diurnal LST estimates (March 13, 2016–March 19, 2016) showed good agreement with the ground-measured LSTs from three sites in northwest China, with R ranging from 0.97 to 0.98 and MAE ranging from 1.86 K to 3.17 K. The predictions from three data sources (i.e., Landsat, MODIS, and CLM) reflected the spatial variability and diurnal temporal dynamics in LST well under all weather conditions, and were much better than the predictions obtained from two data sources (i.e., Landsat and CLM) across all the metrics. In the Landsat-scale evaluation, the temporal normalization showed a positive but limited improvement on the fusion accuracy, while in the site-scale evaluation, this improvement was somewhat increased. Overall, the developed data fusion framework makes full use of the complementary information of satellite-based and LSM-based LST datasets in accuracy, spatio-temporal resolution, and continuity, as well as extending the condition of spatio-temporal LST fusion. Benefiting from the parsimony in other ancillary data (i.e., NDVI, DEM, ground-based LST, etc.), the developed integrated temperature fusion framework expands the transferability to other similar areas for producing all-weather LST estimates. The mechanism of the proposed integrated fusion approach also advances the capability of fusing additional multi-source data (e.g., GF-5, FY-4A, MTSAT-2, MSG SEVIRI) to generate LST estimates of a finer spatio-temporal resolution. Moreover, although the proposed method was developed to generate LST estimates, it could also be transferred to other ecological or environmental parameters, such as evapotranspiration or soil moisture. In our future work, it is anticipated to develop strategies for obtaining more accurate LSM-simulated LST, conduct angular normalization for satellite-observed LST, and design approaches to counter land-cover change in the fusion framework.

**Declaration of competing interest**

The authors declare no conflict of interest.

**Acknowledgements**

This study was supported by the National Key Research and Development Program of China (2019YFB2102900). The reviewers' and editors' comments were highly appreciated. We would like to thank NASA, the USGS, and the Tibetan Plateau Data Center (TPDC) for proving the satellite data, atmospheric forcing datasets, and the in-situ data. We acknowledge Professor Di Long from Tsinghua University for the discussion on the MODIS/Landsat overpass time problem. We acknowledge Professor Kun Yang from Tsinghua University for giving advice on CMFD processing. We also gratefully thank Ph.D. candidate Falu Hong from Nanjing University for the fruitful discussion on the DTC model.

**References**

Abowarda, A.S., Bai, L., Zhang, C., Long, D., Li, X., Huang, Q., & Sun, Z. (2021). Generating surface soil moisture at 30 m spatial resolution using both data fusion and machine learning toward better water resources management at the field scale. *Remote Sensing of Environment, 255*, 112301
Anderson, M.C., Allen, R.G., Morse, A., & Kustas, W.P. (2012). Use of Landsat thermal imagery in monitoring evapotranspiration and managing water resources. *Remote Sensing of Environment, 122*, 50-65

Bei, X., Yao, Y., Zhang, L., Lin, Y., Liu, S., Jia, K., Zhang, X., Shang, K., Yang, J., & Chen, X. (2020). Estimation of daily terrestrial latent heat flux with high spatial resolution from MODIS and Chinese GF-1 data. *Sensors, 20*, 2811

Belgiu, M., & Stein, A. (2019). Spatiotemporal image fusion in remote sensing. *Remote Sensing, 11*, 818

Bojinski, S., Verstraete, M., Peterson, T.C., Richter, C., Simmons, A., & Zemp, M. (2014). The concept of essential climate variables in support of climate research, applications, and policy. *Bulletin of the American Meteorological Society, 95*, 1431-1443

Cao, B., Liu, Q., Du, Y., Roujean, J.-L., Gastellu-Etchegorry, J.-P., Trigo, I.F., Zhan, W., Yu, Y., Cheng, J., & Jacob, F. (2019). A review of earth surface thermal radiation directionality observing and modeling: Historical development, current status and perspectives. *Remote Sensing of Environment, 232*, 111304

Cao, B., Roujean, J.-L., Gastellu-Etchegorry, J.-P., Liu, Q., Du, Y., Lagouarde, J.-P., Huang, H., Li, H., Bian, Z., & Hu, T. (2021). A general framework of kernel-driven modeling in the thermal infrared domain. *Remote Sensing of Environment, 252*, 112157

Chen, J., Wen, J., Tian, H., Zhang, T., Yang, X., Jia, D., & Lai, X. (2018). A Study of Soil Thermal and Hydraulic Properties and Parameterizations for CLM in the SRYR. *Journal of Geophysical Research: Atmospheres, 123*, 8487-8499

Cheng, Q., Liu, H., Shen, H., Wu, P., & Zhang, L. (2017). A spatial and temporal nonlocal filter-based data fusion method. *IEEE Transactions on Geoscience and Remote Sensing, 55*, 4476-4488

Cornette, W.M., & Shanks, J.G. (1993). Impact of cirrus clouds on remote sensing of surface temperatures. In, *Passive Infrared Remote Sensing of Clouds and the Atmosphere* (pp. 252-263): International Society for Optics and Photonics

Deng, M., Meng, X., Lyv, Y., Zhao, L., Li, Z., Hu, Z., & Jing, H. (2020). Comparison of soil water and heat transfer modeling over the Tibetan Plateau using two Community Land Surface Model (CLM) versions. *Journal of Advances in Modeling Earth Systems, 12*, e2020MS002189

Duan, S.-B., Li, Z.-L., & Leng, P. (2017). A framework for the retrieval of all-weather land surface temperature at a high spatial resolution from polar-orbiting thermal infrared and passive microwave data. *Remote Sensing of Environment, 195*, 107-117

Duan, S.-B., Li, Z.-L., Tang, B.-H., Wu, H., & Tang, R. (2014). Generation of a time-consistent land surface temperature product from MODIS data. *Remote Sensing of Environment, 140*, 339-349

Duan, S.-B., Li, Z.-L., Wang, N., Wu, H., & Tang, B.-H. (2012). Evaluation of six land-surface diurnal temperature cycle models using clear-sky in situ and satellite data. *Remote Sensing of Environment, 124*, 15-25

Fu, P., Xie, Y., Weng, Q., Myint, S., Meacham-Hensold, K., & Bernacchi, C. (2019). A physical model-based method for retrieving urban land surface temperatures under cloudy conditions. *Remote Sensing of Environment, 230*, 111191

Gan, W., Shen, H., Zhang, L., & Gong, W. (2014). Normalization of medium-resolution NDVI by the use of coarser reference data: Method and evaluation. *International journal of remote sensing, 35*, 7400-7429

Gao, F., Masek, J., Schwaller, M., & Hall, F. (2006). On the blending of the Landsat and MODIS surface reflectance: Predicting daily Landsat surface reflectance. *IEEE Transactions on Geoscience and Remote Sensing, 44*, 2207-2218

Gao, M., Chen, F., Shen, H., Barlage, M., Li, H., Tan, Z., & Zhang, L. (2019). Efficacy of possible

strategies to mitigate the urban heat island based on urbanized high-resolution land data assimilation system (u-HRLDAS). *Journal of the Meteorological Society of Japan. Ser. II*

Göttsche, F.-M., & Olesen, F.-S. (2009). Modelling the effect of optical thickness on diurnal cycles of land surface temperature. *Remote Sensing of Environment, 113*, 2306-2316

Hansen, J., Ruedy, R., Sato, M., & Lo, K. (2010). Global surface temperature change. *Reviews of Geophysics, 48*

He, J., Yang, K., Tang, W., Lu, H., Qin, J., Chen, Y., & Li, X. (2020). The first high-resolution meteorological forcing dataset for land process studies over China. *Scientific data, 7*, 1-11

Hong, F., Zhan, W., Göttsche, F.-M., Liu, Z., Zhou, J., Huang, F., Lai, J., & Li, M. (2018). Comprehensive assessment of four-parameter diurnal land surface temperature cycle models under clear-sky. *ISPRS Journal of Photogrammetry and Remote Sensing, 142*, 190-204

Huang, B., & Zhang, H. (2014). Spatio-temporal reflectance fusion via unmixing: Accounting for both phenological and land-cover changes. *International journal of remote sensing, 35*, 6213-6233

Huang, F., Zhan, W., Duan, S.-B., Ju, W., & Quan, J. (2014). A generic framework for modeling diurnal land surface temperatures with remotely sensed thermal observations under clear sky. *Remote Sensing of Environment, 150*, 140-151

Hutengs, C., & Vohland, M. (2016). Downscaling land surface temperatures at regional scales with random forest regression. *Remote Sensing of Environment, 178*, 127-141

Jia, A., Ma, H., Liang, S., & Wang, D. (2021). Cloudy-sky land surface temperature from VIIRS and MODIS satellite data using a surface energy balance-based method. *Remote Sensing of Environment, 263*, 112566

Jin, M., & Dickinson, R.E. (1999). Interpolation of surface radiative temperature measured from polar orbiting satellites to a diurnal cycle: 1. Without clouds. *Journal of Geophysical Research: Atmospheres, 104*, 2105-2116

Keramitsoglou, I., Kiranoudis, C.T., & Weng, Q. (2013). Downscaling geostationary land surface temperature imagery for urban analysis. *IEEE Geoscience and remote sensing letters, 10*, 1253-1257

Lawrence, D.M., Fisher, R.A., Koven, C.D., Oleson, K.W., Swenson, S.C., Bonan, G., Collier, N., Ghimire, B., van Kampenhout, L., & Kennedy, D. (2019). The Community Land Model version 5: Description of new features, benchmarking, and impact of forcing uncertainty. *Journal of Advances in Modeling Earth Systems, 11*, 4245-4287

Li, B., Liang, S., Liu, X., Ma, H., Chen, Y., Liang, T., & He, T. (2021). Estimation of all-sky 1 km land surface temperature over the conterminous United States. *Remote Sensing of Environment, 266*, 112707

Li, C., Lu, H., Yang, K., Han, M., Wright, J.S., Chen, Y., Yu, L., Xu, S., Huang, X., & Gong, W. (2018). The evaluation of SMAP enhanced soil moisture products using high-resolution model simulations and in-situ observations on the Tibetan Plateau. *Remote Sensing, 10*, 535

Li, Z.-L., Tang, B.-H., Wu, H., Ren, H., Yan, G., & Wan, Z. (2013). Satellite-derived land surface temperature: Current status and perspectives. *Remote Sensing of Environment, 131*, 14-37

Liu, H., & Weng, Q. (2012). Enhancing temporal resolution of satellite imagery for public health studies: A case study of West Nile Virus outbreak in Los Angeles in 2007. *Remote Sensing of Environment, 117*, 57-71

Liu, S., Li, X., Xu, Z., Che, T., Xiao, Q., Ma, M., Liu, Q., Jin, R., Guo, J., & Wang, L. (2018). The Heihe

Integrated Observatory Network: A basin-scale land surface processes observatory in China. *Vadose Zone Journal, 17*, 1-21

Liu, X., Deng, C., Chanussot, J., Hong, D., & Zhao, B. (2019a). Stfnet: A two-stream convolutional neural network for spatiotemporal image fusion. *IEEE Transactions on Geoscience and Remote Sensing, 57*, 6552-6564

Liu, Z., Zhan, W., Lai, J., Hong, F., Quan, J., Bechtel, B., Huang, F., & Zou, Z. (2019b). Balancing prediction accuracy and generalization ability: A hybrid framework for modelling the annual dynamics of satellite-derived land surface temperatures. *ISPRS Journal of Photogrammetry and Remote Sensing, 151*, 189-206

Lombardozzi, D.L., Lu, Y., Lawrence, P.J., Lawrence, D.M., Swenson, S., Oleson, K.W., Wieder, W.R., & Ainsworth, E.A. (2020). Simulating agriculture in the Community Land Model version 5. *Journal of Geophysical Research: Biogeosciences, 125*, e2019JG005529

Long, D., Yan, L., Bai, L., Zhang, C., Li, X., Lei, H., Yang, H., Tian, F., Zeng, C., & Meng, X. (2020). Generation of MODIS-like land surface temperatures under all-weather conditions based on a data fusion approach. *Remote Sensing of Environment, 246*, 111863

Lu, L., Venus, V., Skidmore, A., Wang, T., & Luo, G. (2011). Estimating land-surface temperature under clouds using MSG/SEVIRI observations. *International journal of applied earth observation and geoinformation, 13*, 265-276

Ma, X., Jin, J., Zhu, L., & Liu, J. (2021). Evaluating and improving simulations of diurnal variation in land surface temperature with the Community Land Model for the Tibetan Plateau. *PeerJ, 9*, e11040

Oleson, K., & Feddema, J. (2020). Parameterization and surface data improvements and new capabilities for the Community Land Model Urban (CLMU). *Journal of Advances in Modeling Earth Systems, 12*, e2018MS001586

Oleson, K.W., Bonan, G.B., Feddema, J., Vertenstein, M., & Kluzek, E. (2010). Technical description of an urban parameterization for the Community Land Model (CLMU). *NCAR, Boulder*

Pede, T., & Mountrakis, G. (2018). An empirical comparison of interpolation methods for MODIS 8-day land surface temperature composites across the conterminous Unites States. *ISPRS Journal of Photogrammetry and Remote Sensing, 142*, 137-150

Quan, J., Zhan, W., Ma, T., Du, Y., Guo, Z., & Qin, B. (2018). An integrated model for generating hourly Landsat-like land surface temperatures over heterogeneous landscapes. *Remote Sensing of Environment, 206*, 403-423

Shen, H., Huang, L., Zhang, L., Wu, P., & Zeng, C. (2016). Long-term and fine-scale satellite monitoring of the urban heat island effect by the fusion of multi-temporal and multi-sensor remote sensed data: A 26-year case study of the city of Wuhan in China. *Remote Sensing of Environment, 172*, 109-125

Shen, H., Li, X., Cheng, Q., Zeng, C., Yang, G., Li, H., & Zhang, L. (2015). Missing information reconstruction of remote sensing data: A technical review. *IEEE Geoscience and Remote Sensing Magazine, 3*, 61-85

Shen, Y., Shen, H., Cheng, Q., & Zhang, L. (2020). Generating Comparable and Fine-Scale Time Series of Summer Land Surface Temperature for Thermal Environment Monitoring. *IEEE Journal of Selected Topics in Applied Earth Observations and Remote Sensing, 14*, 2136-2147

Siemann, A.L., Coccia, G., Pan, M., & Wood, E.F. (2016). Development and analysis of a long-term, global, terrestrial land surface temperature dataset based on HIRS satellite retrievals. *Journal of*

*Climate, 29*, 3589-3606

Song, H., & Huang, B. (2012). Spatiotemporal satellite image fusion through one-pair image learning. *IEEE Transactions on Geoscience and Remote Sensing, 51*, 1883-1896

Song, J., Miller, G.R., Cahill, A.T., Aparecido, L.M.T., & Moore, G.W. (2020). Modeling land surface processes over a mountainous rainforest in Costa Rica using CLM4. 5 and CLM5. *Geoscientific Model Development, 13*, 5147-5173

Survey, U.S.G. (U.S. Geological Survey, 2021). Landsat 8 Collection 2 (C2) Level 2 Science Product (L2SP) Guide. In: U.S. Geological Survey

Swenson, S., & Lawrence, D. (2014). Assessing a dry surface layer-based soil resistance parameterization for the Community Land Model using GRACE and FLUXNET-MTE data. *Journal of Geophysical Research: Atmospheres, 119*, 10,299-210,312

Trigo, I., Boussetta, S., Viterbo, P., Balsamo, G., Beljaars, A., & Sandu, I. (2015). Comparison of model land skin temperature with remotely sensed estimates and assessment of surface-atmosphere coupling. *Journal of Geophysical Research: Atmospheres, 120*, 12,096-012,111

Wan, Z. (2014). New refinements and validation of the collection-6 MODIS land-surface temperature/emissivity product. *Remote Sensing of Environment, 140*, 36-45

Wan, Z., & Dozier, J. (1996). A generalized split-window algorithm for retrieving land-surface temperature from space. *IEEE Transactions on Geoscience and Remote Sensing, 34*, 892-905

Wang, A., Barlage, M., Zeng, X., & Draper, C.S. (2014a). Comparison of land skin temperature from a land model, remote sensing, and in situ measurement. *Journal of Geophysical Research: Atmospheres, 119*, 3093-3106

Wang, A., Yang, Y., Pan, X., Zhang, Y., & Hujie, J. (2021a). Research on land surface temperature downscaling method based on diurnal temperature cycle model deviation coefficient calculation. *Journal of Remote Sensing, 25*, 1735-1748

Wang, D., Chen, Y., Hu, L., Voogt, J.A., Gastellu-Etchegorry, J.-P., & Krayenhoff, E.S. (2021b). Modeling the angular effect of MODIS LST in urban areas: A case study of Toulouse, France. *Remote Sensing of Environment, 257*, 112361

Wang, J., Li, C., & Gong, P. (2015). Adaptively weighted decision fusion in 30 m land-cover mapping with Landsat and MODIS data. *International journal of remote sensing, 36*, 3659-3674

Wang, K., & Liang, S. (2009). Evaluation of ASTER and MODIS land surface temperature and emissivity products using long-term surface longwave radiation observations at SURFRAD sites. *Remote Sensing of Environment, 113*, 1556-1565

Wang, K., Wan, Z., Wang, P., Sparrow, M., Liu, J., Zhou, X., & Haginoya, S. (2005). Estimation of surface long wave radiation and broadband emissivity using Moderate Resolution Imaging Spectroradiometer (MODIS) land surface temperature/emissivity products. *Journal of Geophysical Research: Atmospheres, 110*

Wang, Q., & Atkinson, P.M. (2018). Spatio-temporal fusion for daily Sentinel-2 images. *Remote Sensing of Environment, 204*, 31-42

Wang, T., Shi, J., Ma, Y., Husi, L., Comyn-Platt, E., Ji, D., Zhao, T., & Xiong, C. (2019). Recovering land surface temperature under cloudy skies considering the solar-cloud-satellite geometry: Application to MODIS and Landsat-8 data. *Journal of Geophysical Research: Atmospheres, 124*, 3401-3416

Wang, T., Shi, J., Yan, G., Zhao, T., Ji, D., & Xiong, C. (2014b). Recovering land surface temperature under cloudy skies for potentially deriving surface emitted longwave radiation by fusing MODIS and AMSR-E measurements. In, *2014 IEEE Geoscience and Remote Sensing Symposium* (pp. 1805-1808): IEEE

Weng, Q., Fu, P., & Gao, F. (2014). Generating daily land surface temperature at Landsat resolution by fusing Landsat and MODIS data. *Remote Sensing of Environment, 145*, 55-67

Wu, P., Shen, H., Zhang, L., & Göttsche, F.-M. (2015). Integrated fusion of multi-scale polar-orbiting and geostationary satellite observations for the mapping of high spatial and temporal resolution land surface temperature. *Remote Sensing of Environment, 156*, 169-181

Wu, P., Yin, Z., Zeng, C., Duan, S., Gottsche, F.-M., Ma, X., Li, X., Yang, H., & Shen, H. (2019). Spatially Continuous and High-resolution Land Surface Temperature: A Review of Reconstruction and Spatiotemporal Fusion Techniques. *arXiv preprint arXiv:1909.09316*

Xu, S., & Cheng, J. (2021). A new land surface temperature fusion strategy based on cumulative distribution function matching and multiresolution Kalman filtering. *Remote Sensing of Environment, 254*, 112256

Yang, G., Sun, W., Shen, H., Meng, X., & Li, J. (2019). An integrated method for reconstructing daily MODIS land surface temperature data. *IEEE Journal of Selected Topics in Applied Earth Observations and Remote Sensing, 12*, 1026-1040

Yin, Z., Wu, P., Foody, G.M., Wu, Y., Liu, Z., Du, Y., & Ling, F. (2020). Spatiotemporal fusion of land surface temperature based on a convolutional neural network. *IEEE Transactions on Geoscience and Remote Sensing*

Zakšek, K., & Oštir, K. (2012). Downscaling land surface temperature for urban heat island diurnal cycle analysis. *Remote Sensing of Environment, 117*, 114-124

Zeng, C., Long, D., Shen, H., Wu, P., Cui, Y., & Hong, Y. (2018). A two-step framework for reconstructing remotely sensed land surface temperatures contaminated by cloud. *ISPRS Journal of Photogrammetry and Remote Sensing, 141*, 30-45

Zeng, C., Shen, H., & Zhang, L. (2013). Recovering missing pixels for Landsat ETM+ SLC-off imagery using multi-temporal regression analysis and a regularization method. *Remote Sensing of Environment, 131*, 182-194

Zhan, W., Chen, Y., Zhou, J., Wang, J., Liu, W., Voogt, J., Zhu, X., Quan, J., & Li, J. (2013). Disaggregation of remotely sensed land surface temperature: Literature survey, taxonomy, issues, and caveats. *Remote Sensing of Environment, 131*, 119-139

Zhan, W., Huang, F., Quan, J., Zhu, X., Gao, L., Zhou, J., & Ju, W. (2016). Disaggregation of remotely sensed land surface temperature: A new dynamic methodology. *Journal of Geophysical Research: Atmospheres, 121*, 10,538-510,554

Zhang, X., Zhou, J., Göttsche, F.-M., Zhan, W., Liu, S., & Cao, R. (2019). A method based on temporal component decomposition for estimating 1-km all-weather land surface temperature by merging satellite thermal infrared and passive microwave observations. *IEEE Transactions on Geoscience and Remote Sensing, 57*, 4670-4691

Zhang, X., Zhou, J., Liang, S., & Wang, D. (2021). A practical reanalysis data and thermal infrared remote sensing data merging (RTM) method for reconstruction of a 1-km all-weather land surface temperature. *Remote Sensing of Environment, 260*, 112437

Zhao, W., & Duan, S.-B. (2020). Reconstruction of daytime land surface temperatures under cloud-covered conditions using integrated MODIS/Terra land products and MSG geostationary satellite data. *Remote Sensing of Environment, 247*, 111931

Zhao, W., Wu, H., Yin, G., & Duan, S.-B. (2019). Normalization of the temporal effect on the MODIS

land surface temperature product using random forest regression. *ISPRS Journal of Photogrammetry and Remote Sensing, 152*, 109-118

Zhao, Y., Huang, B., & Song, H. (2018). A robust adaptive spatial and temporal image fusion model for complex land surface changes. *Remote Sensing of Environment, 208*, 42-62

Zhu, L., Zhou, J., Liu, S., Li, M., & Li, G. (2016a). Comparison of diurnal temperature cycle model and polynomial regression technique in temporal normalization of airborne land surface temperature. In, *2016 IEEE International Geoscience and Remote Sensing Symposium (IGARSS)* (pp. 4309-4312): IEEE

Zhu, X., Chen, J., Gao, F., Chen, X., & Masek, J.G. (2010). An enhanced spatial and temporal adaptive reflectance fusion model for complex heterogeneous regions. *Remote Sensing of Environment, 114*, 2610-2623

Zhu, X., Helmer, E.H., Gao, F., Liu, D., Chen, J., & Lefsky, M.A. (2016b). A flexible spatiotemporal method for fusing satellite images with different resolutions. *Remote Sensing of Environment, 172*, 165-177

Zhu, X., Song, X., Leng, P., Li, X., Gao, L., Guo, D., & Cai, S. (2021). A Framework for Generating High Spatiotemporal Resolution Land Surface Temperature in Heterogeneous Areas. *Remote Sensing, 13*, 3885